\documentclass[12pt]{article}
\usepackage{graphicx}
\usepackage[utf8]{inputenc}
\usepackage{appendix}

\usepackage{amsmath,amssymb,bbm,color}
\usepackage{amsbsy}
\usepackage{euscript}
\usepackage{graphicx}
\usepackage{hyperref}
\usepackage{cite}
\usepackage{enumerate}
\numberwithin{equation}{section}

\usepackage[]{changes} 

\newcommand{\order}[1]{${\cal O}(#1)$}
\newcommand{\Aver}[1]{\langle #1 \rangle}

\newcommand{\kkmceeFive}{{\tt KKMCee 5.00.2}}

\newcommand{\kkmcee}{{\tt KKMCee}}
\newcommand{\kkmchh}{{\tt KKMChh}}
\newcommand{\kkmc}{{\tt KKMC}}
\newcommand{\foam}{{\tt FOAM}}
\newcommand{\kkeefoam}{{\tt KKeeFoam}}
\newcommand{\dizet}{{\tt DIZET}}
\newcommand{\tauola}{{\tt TAUOLA}}
\newcommand{\photos}{{\tt PHOTOS}}

\newcommand{\hepmc}{{\tt HEPMC}}
\newcommand{\photospp}{{\tt Photospp}}

\begin{document}

\begin{titlepage}
\begin{flushright}
\bf IFJPAN-IV-2022-6, BU-HEPP-22-02, MCnet-22
\end{flushright}
\vspace{5mm}
\begin{center}
    {\Large\bf Multi-photon Monte Carlo event generator {\tt KKMCee}
    for lepton and quark pair production
    in lepton colliders$^{\star}$}
\end{center}

\vspace{1mm}
\begin{center}
  {\bf   S. Jadach$^{a}$,}
  {\bf   B.F.L. Ward$^{b}$,}
  {\bf   Z. W\c{a}s$^{a}$,}
  {\bf   S.A. Yost$^{c}$}
  {\em and}
  {\bf   A. Siodmok$^{d}$}
\\
\vspace{1mm}
\vspace{1mm}
{\em $^a$Institute of Nuclear Physics, Polish Academy of Sciences,\\
  ul.\ Radzikowskiego 152, 31-342 Krak\'ow, Poland},\\
{\em $^b$Department of Physics,
   Baylor University, Waco, TX 76798-7316, USA,}\\
{\em $^c$The Citadel, Charleston, SC, USA }\\
{\em $^d$Institute of Applied Computer Science, Jagiellonian University,\\
  ul. prof. Stanisława Łojasiewicza 11,
  30-348 Krak\'ow, Poland}
\end{center}

\vspace{2mm}
\begin{abstract}
\noindent
 We present the {\tt KKMCee 5.00.2} Monte Carlo event generator for lepton and quark pair
 production for the high energy electron-positron annihilation process.
 It is still the most sophisticated event generator for such processes.
 Its entire source code is re-written in the modern C++ language.
 It reproduces all features of the older \kkmc\ code in Fortran 77.
 However, a number of improvements in the Monte Carlo algorithm are also implemented.
 Most importantly, it is intended to be a starting point for the future improvements,
 which will be mandatory for the future high precision lepton collider projects.
 As in the older version, in addition to higher order QED corrections,
 it includes so-called \order{\alpha^{1.5}} genuine weak corrections
 using a  version of the classic {\tt DIZET} library
 and polarized $\tau$ decays using {\tt TAUOLA} program.
 Both {\tt DIZET} and {\tt TAUOLA} external libraries are still in Fortran 77.
 In addition, a {\tt HEPMC3} interface to other MC programs, 
 like parton showers and detector simulation,
 replaces the older {\tt HepEvt} interface. The
 {\tt HEPMC3} interface is also exploited in the implementation of
 the additional photon final state emissions in $\tau$ decays
 using an external {\tt PHOTOS} library rewritten in C.
\end{abstract}

\vspace{20mm}
\footnoterule
\noindent
{\footnotesize
$^{\star}$%
This work is partly supported by the National Science Centre, 
Poland grants No. 2019/34/E/ST2/00457 and 2017/27/B/ST2/01391,
by the EU Horizon 2020 research and innovation programme 
under grant agreement No. 951754
and the CERN FCC Design Study Programme. 
The work of AS was also funded by 
the Priority Research Area Digiworld under the program Excellence Initiative
– Research University at the Jagiellonian University in Cracow.
The work of SAY was partly supported by a grant from The Citadel Foundation.
}
\begin{flushleft}
{\bf 
  IFJPAN-IV-2022-6, BU-HEPP-22-02, MCnet-22\\
  April 2022}
\end{flushleft}

\end{titlepage}

\noindent{\bf \large PROGRAM SUMMARY}
\vspace{10pt}

\noindent{\bf Manuscript title:} Monte Carlo event generator {\tt KKMCee} Technical
and Physics Documentation

\noindent{\bf Authors:} \- S. Jadach and B.F.L Ward, Z. W\c as, S.A. Yost and A. Si\'odmok.

\noindent{\bf Program title:} \- \kkmceeFive

\noindent{\bf Licensing provisions:} \- 
GPL-3.0

\noindent{\bf Programming languages:} \- {\tt C++, FORTRAN77 }

\noindent{\bf Operating system(s) for which the program has been designed:} 
\- Linux

\noindent{\bf RAM required to execute with typical data:} \- $<$100MB

\noindent{\bf Has the code been vectorised or parallelized?:} \- No

\noindent{\bf Number of processors used:} \- 1

\noindent{\bf Supplementary material:} \- None


\noindent{\bf CPC Library Classification:} 11.2 - Phase Space and Event Simulation

\noindent{\bf External routines/libraries used:} 
 {\tt CERN ROOT library}, {\tt PHOTOS}, {\tt HepMC v.3.0},

\noindent{\bf CPC Program Library subprograms used:} 
 {\tt TAUOLA, PHOTOS, FOAM, HEPMC3 }

\noindent{\bf Nature of the problem:}\\
Fermion pair production is and will be used as an important  data  source
for  precise  tests  of  the  standard  electroweak  theory  
at a high luminosity future circular collider near the $Z$ resonance and above and/or
at  the future linear lepton  colliders  of  higher energies than those at LEP.
The QED  corrections  to  fermion  pair  production (especially $\tau$ leptons) 
have to be known to at least second order,
including  spin  polarization  effects, with 4-5 digit precision. 
The  Standard  Model  predictions  at  the  sub-permille precision level, 
taking into account multiple emission of photons for realistic experimental acceptances, 
can only be obtained using a Monte Carlo event generator.
The realistic and precise simulation of $\tau$ lepton decays taking into account
spin effects is an indispensable ingredient in the Monte Carlo event generator
for the fermion pair production process.

\noindent{\bf Solution method:}\\
Monte  Carlo  methods  are  used  to  simulate  most  of  
the  two-fermion final-state processes in $e^+e^-$ collisions in 
the presence of multiphoton initial and final state radiation.  
The multiphoton effects are described in the framework 
of coherent exclusive exponentiation (CEEX) extending/upgrading 
the older Yennie–Frautschi–Suura exclusive exponentiation (EEX) scheme.
CEEX treats  correctly  to  infinite  order  not  only  infrared  cancellations  
but  also  QED  interferences, including suppression of initial-final
state interferences for narrow resonances.  
The matrix element according to the older YFS exponentiation 
is also implemented for the testing purpose.
For $\tau$ leptons, the appropriate simulation of a very rich spectrum 
of the decays is included.  
Beam polarization and spin effects, 
both longitudinal and transverse, in tau decays are properly taken into account. 
Gaussian beam spread and an arbitrary spectrum of the beamstrahlung 
are also optionally simulated.
The present version of the program is rewritten to C++ but in many respects
corresponds to its FORTRAN predecessor {\tt KKMC} v.4.13 [1] 
with later minor modifications in v. 4.32 [2].  

\noindent{\bf Restrictions:}\\
In the present version, electron (Bhabha) and $t$-quark final states are not included.  
(It is planned for a future version.)
Third-order  QED  corrections  in the leading-logarithmic approximation are included 
only in the auxiliary older YFS/EEX matrix element.
The electroweak corrections should not be trusted above the $t$-quark threshold.
The total cross section for light quarks for $\sqrt{s} <10$ GeV 
(including narrow resonances) requires an improvement using experimental data.
The program does not provide any handles for the beyond the Standard Model (BSM)
physics, for instance in the Born $Z$ boson couplings 
or in the electroweak (EW) formfactors.

\noindent{\bf Running time:}\\
Depends on the CMS energy, final fermion type, 
upper phase space limit of the photon energy,
and whether variable weight events or WT=1 events are generated.
On a PC/Linux with a 2.2 GHz processor,
producing 100k variable weight events at $\sqrt{s}=M_Z$ takes 25 sec. of CPU time
for  $\mu$-pairs and 30 sec. for $\tau$-pairs including decays.
At $\sqrt{s}=189$GeV 100k events with WT=1 costs 1200 sec. for $\mu$-pairs
and 830 sec. for $\tau$-pairs (less hard photons).

\vspace{1mm}
\noindent{\bf References: } \\
$[1]$ S. Jadach, B. F. L. Ward and Z. Was, 
{\em ``The Precision Monte Carlo event generator KK
for two fermion final states in $e^+e^-$ collisions''}, 
Comput. Phys. Commun. \textbf{130} (2000) 260.\\
$[2]$  A. Arbuzov, S. Jadach, Z. Was, B.F.L. Ward and S.A. Yost,
{\em ``The Monte Carlo Program KKMC, for the Lepton or Quark Pair Production 
at LEP/SLC Energies -- Updates of electroweak calculations''},
Comput. Phys. Commun. \textbf{260} (2021) 107734.

\newpage
\tableofcontents
\newpage

\newpage
\section{Introduction}
The \kkmceeFive\ program presented in this work
is the Monte Carlo (MC) event generator for the fermion 
pair production process in the electron-positron annihilation
accompanied with multiple photons:\\
\centerline{$e^-e^+ \to f\bar{f}+ n\gamma$,~~
            $f=\mu,\tau,\nu,u,d,s,c,b$, $n=0,1,2,\ldots,\infty$.}
            
The first \kkmc\ version 4.13~\cite{Jadach:1999vf} was released in the year 2000%
\footnote{The prototype version of \kkmc\ was presented 
already in a CERN TH seminar in October 1998
\centerline{\href{https://nz42.ifj.edu.pl/_media/user/jadach/main/talks/ceex-talk-5nov98.pdf}%
{\tt https://nz42.ifj.edu.pl/\_media/user/jadach/main/talks/ceex-talk-5nov98.pdf}}}.
The publication~\cite{Jadach:1999vf} contains the very detailed (80 pages) description of the Monte Carlo algorithm,
of the Fortran 77 code and it is the user's guide (manual).
The two other Refs.~\cite{Jadach:1998jb,Jadach:2000ir} 
published at about the same time
described the physics content of \kkmc, in particular the
latter  formulated  the QED coherent exclusive exponentiation
(CEEX) scheme, including soft (and hard) photon resummation.
\kkmc\ has replaced two earlier similar MC's, 
{\tt KORALZ}~\cite{Jadach:1991ws} and {\tt KORALB}~\cite{Jadach:1994ps},
which continue to be the source of useful benchmarks.
It is not applicable for the $e^-e^+ \to e^-e^+$ Bhabha process.
From the very beginning \kkmc\ was interfaced
with the \tauola\ and \photos\ programs~\cite{Jadach:1993hs,Barberio:1990ms}
for simulating decays of polarized $\tau$'s 
and with the library \dizet~\cite{Bardin:1989tq} of the complete \order{\alpha^1} 
electroweak (EW) corrections.
The complete spin polarization (longitudinal and transverse)
implementation in \kkmc\ was outlined in Ref.~\cite{Jadach:1998wp}.
Note that most of the description of the input data and overview of physics
in Ref.~\cite{Jadach:1999vf} remains actual
and is not repeated in this paper.

In this work we present a new version 
of \kkmc\ renamed as \kkmceeFive, which is written entirely in C++.
For the moment \kkmcee\ is functionally not much different from
the latest versions in Fortran 77 (F77), hence the natural question is:
{\em Why was rewriting \kkmc\ from F77 to C++ worth pursuing?}

One important reason is that \kkmc\ is widely used in the data analysis
of all existing electron colliders like BES or BELLE
and in studies on the future electron colliders like FCCee, CLIC, ILC,
within a software environment which is mostly programmed in C++.
It will therefore be easier to interface \kkmcee\ with the software
of current and future experiments.
Many past theoretical studies analysing events from \kkmc\ were already
done using elaborate programs written entirely in C++. 
It will also be easier to develop more future theoretical 
studies with \kkmcee\  C++.

Another even more important reason is that,
in spite of the fact the \kkmc\
with the QED coherent exclusive exponentiation (CEEX) matrix element
is still the most sophisticated Monte Carlo program for the fermion
pair production in electron colliders,
it requires many improvements in order to meet the very high precision specifications
of the future electron colliders, for instance the TeraZ option of FCCee,
see Refs.~\cite{FCC:2018evy,Blondel:2018mad,Jadach:2019bye}.
Developing improved versions of the CEEX matrix element,
with even higher orders, more subleading corrections, 
and more versatility for porting it to other processes
definitely will be easier using a more sophisticated programming language.
A number of pending improvements in the Monte Carlo generation
of the multiphoton phase space will be also easier to realize in C++.
Last but not the least, it is inevitable that the next versions
of the auxiliary program libraries 
like \tauola\ for $\tau$ lepton decay, the
next libraries of the electroweak corrections,
and parton shower MCs such as 
{\tt Herwig}~\cite{Bahr:2008pv,Bellm:2015jjp,Bellm:2019zci},
{\tt Pythia}~\cite{Sjostrand:2007gs,Sjostrand:2006za} or 
{\tt Sherpa}~\cite{Gleisberg:2008ta,Sherpa:2019gpd} 
for hadronizing final state quark pairs,
will be, or are already, all in C++. 
In fact, \photos\ is already rewritten to C \cite{Davidson:2010ew}
and is interfaced to the present \kkmcee.
For \tauola\ the first preparatory step in this direction
is documented in \cite{Antropov:2019ald} (see also \cite{Banerjee:2021rtn})
and the on-going effort on the \tauola\ improvements is expected 
to rely on fits to Belle II data. 

Note also that the \kkmc\ project was split recently
into the \kkmcee\ branch for lepton colliders presented here
and the \kkmchh\ branch for hadron colliders.
An unpublished, but quite well tested,
\kkmchh\ version written entirely in C++ already exists
and is in use for studies related to LHC data.

Since many features of the present \kkmcee\ version are
the same or very similar as in the original 
version of Ref.~\cite{Jadach:1999vf},
their detailed description will be not repeated.
Because of that the present documentation is not self contained
and requires occasionally consulting%
\footnote{In particular tables of input parameters 
  are described in Tables 2-5 in Ref.~\cite{Jadach:1999vf}.
  However, they are also documented in 
  comments in the {\tt SRCee/KKMCee\_defaults} file.}
Ref.~\cite{Jadach:1999vf}.
We shall concentrate on the differences with the original \kkmc\ 
and in particular on some improvements implemented in the new \kkmcee.
More precisely the present version is inheriting 
all features of the F77 version except the hadronization interface
for the final massive quark pair using the parton shower MC
and the {\tt RRes} package%
\footnote{ {\tt RRes} package was provided by Maarten Boonekamp in 2001, 
  hence is not quoted/documented in Ref.~\cite{Jadach:1999vf}. 
  In case the quark pair mass is below $\sim 10$GeV it implements the experimental
  $e^+e^-\to q\bar{q}$ cross section including most of narrow resonances
  and simulates resonance decays using either the {\tt Pythia 6}
  Monte Carlo or its own subprograms.
}
for low energy quark pairs.
We hope to recover this feature in the future updates.

On the other hand, there are many improvements of the MC algorithm,
structure of the program, and program performance in terms of CPU time.
More details are in the following sections -- here
let us summarize briefly on all these improvement:
\begin{itemize}
\item 
\foam\ takes a central role in the MC algorithm.
It generates the energy spread of both beams according to an arbitrary distribution,
total energy of the initial state radiation (ISR),
type of final state fermion and the scattering angle $\theta_f$ of the final fermion.
This leads to simplification of the MC algorithm and the code.
\item
Thanks to generation of $\cos\theta_f$ by \foam\ (instead of a flat distribution)
the CPU consumption for $WT=1$ events is reduced by a factor of 2 and
for the electron neutrino channel by up to a factor of 20.
\item
The auxiliary program {\tt KKsem} for comparing \kkmc\ results
with semi-analytical formulas is now replaced with the
much more powerful {\tt KKfoam} tool based on \foam.
\item
The persistency mechanism is now implemented (see Sect. 2.2) 
using {\tt ROOT}, providing more flexibility in the generation and analysis
of the MC events.
The \kkmc\ code is now slimmer,
because {\tt ROOT} provides many services like random numbers, Lorentz kinematics,
histogramming, and graphics, formerly handled by the \kkmc\ code.
\item
The handling of input parameters and of the entire
data base of all physics parameters and steering variables is now more
transparent and systematic thanks to use of
an object of the dedicated C++ class. 
\item
The most advanced and sizable CEEX type QED matrix element code 
is now more compact and transparent thanks to introduction of
the auxiliary C++ classes.
\item
The \hepmc3 event record is instrumental for
interfacing to the latest version of PHOTOS and  will facilitate the interfacing
of \kkmcee\ to parton modern shower MCs and to detector simulation 
of the collider experiments.
\end{itemize}

Before we describe \kkmcee\ in a more detail,
let us summarize briefly the evolution of the Fortran 77 code of \kkmc\
since it was first published in 2000 until today.

\subsection{The evolution of the \kkmc\ code}
The  production version  4.16 (October 2001) of \kkmc\   
was a ``workhorse'' in the final data analysis of 
all four LEP collaborations.
It featured an improved matrix element 
for the Standard Model neutrino pair final state
and the {\tt RRes} module for the decay
of the off-shell $\gamma^*$ into narrow~resonances%
\footnote{ {\tt RRes} is not included in the present version.}.

Over the next years, two improved public versions of \kkmc\
were made available:\\
(i) The development version 4.19 (Sept. 2002),  
with added C++ wrappers,
further improvements of the  matrix element for $\nu\bar{\nu}$ final states
and {\tt RRes} for low energy resonances.
The complete second order subleading QED corrections were added
according to Ref.~\cite{Jadach:2001jx}.
The beam spread due to collinear beamstrahlung was implemented
for the NLC/ILC studies using the 
fortran version of the \foam\ program~\cite{Jadach:1999sf}.
\\
(ii) The development version  4.22 (June 2013) included
a possibility of $\mu^-\mu^+$ and $q\bar{q}$ beams
(instead of $e^-e^+$) at the fixed energy.
Optionally, collinear PDFs for $q\bar{q}$ beams, 
replacing beamstrahlung distributions,
were available as a patch in the source code (a temporary solution).
In addition a lot of technical improvements were done,
for instance the automake/autotools system was introduced.
The above versions of \kkmc\ are available from the archive web page:\\
\centerline{\href{http://jadach.web.cern.ch/jadach/KKindex.html}%
{\tt http://jadach.web.cern.ch/jadach/KKindex.html}}.

In 2017, the development of the \kkmc\ was split
into \kkmcee\ for lepton colliders and \kkmchh\ for hadron colliders.
The first Fortran 77 version of the \kkmcee\ is numbered as 4.24.
It is well tested under modern linux systems (mainly on Centos  and Ubuntu),
its implementation of beamstrahlung was improved and simplified
(insertions for $q\bar{q}$ beams are removed).

The public source code of the latest
F77 version (Oct. 2021) of the \kkmcee\ is numbered as 4.32.
It features a number of important improvements:
the electroweak  library \dizet\ was upgraded to version 6.45, 
see Ref.~\cite{Arbuzov:2020coe},
\tauola\ was upgraded to the version 3.1 of Ref.~\cite{Chrzaszcz:2016fte},
MC events in the Les Houches format are provided.
This last F77 version resides in the public {\tt github.com} repository:\\
\centerline{\href{https://github.com/KrakowHEPSoft/KKMCee/releases/tag/v4.32.01}%
{\tt https://github.com/KrakowHEPSoft/KKMCee/releases/tag/v4.32.01}}.
The above F77 version will not be developed any further --
only minor corrections are envisaged.
The present C++ version 5.00 replaces it and will be developed
in the future.
It is now available at\\
\centerline{\href{https://github.com/KrakowHEPSoft/KKMCee/releases/tag/v5.00.00}%
{\tt https://github.com/KrakowHEPSoft/KKMCee/releases/tag/v5.00.00}}.

\vspace{2mm}
The \kkmc\ distributions' directories from its early versions
always included a semi-analytical program {\tt KKsem},
which provided the total cross section and charge asymmetry
as a function of the cutoff on the total photon energy,
obtained using partial analytical integration over the phase space
and partial numerical 3-dimensional non-MC integration.
The comparison of \kkmc\ results with {\tt KKsem} 
for these semi-inclusive observables provided
a very useful crosscheck of the correctness of the \kkmc\ event generator
and an estimate of its overall precision.
Such a comparison, Ref.~\cite{Jadach:2000ir}, led to an
estimate of the generic \kkmc\ precision of  $\sim 0.2\%$ --
good enough for LEP experiments.
The {\tt KKsem} program includes initial state radiation (ISR)
and final state radiation (FSR) but 
not the initial-final state interference (IFI).
For the evaluation of IFI component ambiguities 
an additional comparison~\cite{Jadach:2000ir} with 
the semi-analytical code {\tt ZFITTER}~\cite{Bardin:1999yd} was used.
With the advent of the FCCee related studies, it became urgent to re-examine
the question of the  \kkmc\ precision, especially for the TeraZ option
near the $Z$ resonance, where the experimental precision for the charge asymmetry
may reach the level of $10^{-5}$.
For this task, in Ref.~\cite{Jadach:2018lwm},
a new analytical integration including the IFI contribution was prepared 
and a new program \kkeefoam\ written in C++,
with a 5-dimensional numerical integration using \foam\ \cite{Jadach:2002kn}, was created.
Using \kkeefoam\ a new precision estimate of the \kkmc\
charge asymmetry and integrated cross section for the CEEX matrix element
was established at the level close to $10^{-4}$.
The \kkeefoam\ program is not included in the public \kkmcee\ version 4.32
on {\tt github.com}, but
it is now included in the C++ version of \kkmcee.

As already said, the most important aim of the C++ version is to facilitate future
developments of the \kkmcee.
For more details see Sect.~\ref{sec:prospects}.

\section{Structure of the program}
The source code of \kkmc\ was originally written entirely in Fortran 77 (F77)
simply because at the time%
\footnote{The first version of \kkmc\ was tested in 1998.}
the C++ programming language
was not yet well-established in scientific computing.
However, the \kkmc\ source code in F77 was organized from the very beginning
in such a manner that it would be easy to translate it into C++ in the future.
In particular the F77 code was organized into modules (called pseudo-classes)
resembling C++ classes.
Each module has only one common block included in all subroutines belonging
to this module. 
Different modules communicate each with other only through special
functions, the so-called ``setters'' and ``getters''.
This structure of the F77 code is now partly reflected in the new C++ code.

The core \kkmcee\ Monte Carlo event generator
is not completely standalone - it relies on a certain
programming environment, which resides partly in the operating system
and partly in the local distribution directory.
The system-resident tools/libraries are automake/autotools,
CERN {\tt ROOT}~\cite{Brun:1997pa} library version 6, 
\hepmc3 library \cite{Buckley:2019xhk}
and \photospp~\cite{Davidson:2010ew}.
Other local software libraries are \foam, {\tt MCevelop} and auxiliary
physics libraries are \tauola\ and the electroweak library \dizet.
Their documentation is available respectively
in~\cite{Jadach:1993hs,Bardin:1989tq}.

Let us now go through the main classes of the core MC generator code
and explain their components and role.

\subsection{Top level {\tt KKee2f} class}
This class is the central hub of the event generator.
It contains pointers to ``component objects'' of all other classes.
It allocates, initializes and runs all these objects.
Its data members include only the key input parameters 
and event parameters -- all other variables are encapsulated 
as data members in the component objects.
Let us list all component objects of the {\tt KKee2f} class:
\begin{verbatim}
 KKdbase *DB;                     // Database, input parameters
 TWtMon   *m_WtMainMonit;         // Monitoring WtMain
 KKborn   *m_BornDist;            // Born matrix element
 KKdizet  *m_EWtabs;              // EW formfactors
 KKevent  *m_Event;               // MC event ISR+FSR in KKMC format
 GenEvent *m_Hvent;               // MC event in HEPMC3 format
 KKarLud  *m_GenISR;              // ISR YFS multiphoton generator
 KKarFin  *m_GenFSR;              // FSR YFS multiphoton generator
 KKqed3   *m_QED3;                // EEX matrix element
 KKceex   *m_GPS;                 // CEEX matrix element
 KKbvir   *m_BVR;                 // Library of virtual corrections
 TauPair  *m_TauGen;              // Interface to TAUOLA+PHOTOS
 HepFace  *m_HEPMC;               // Interface to HEPMC3 event
\end{verbatim}
Their role and the corresponding classes will be characterized in the following.
In addition,
three important member objects (pointers) in the {\tt KKee2f} class
are inherited from the base class {\tt TMCgen}
of the {\tt MCdevelop} library of tools for building
Monte Carlo event generators, see Ref.~\cite{Slawinska:2010jn} for its documentation:
\begin{verbatim}
  TRandom  *f_RNgen;            //  External RN event generator
  TFOAM    *f_FoamI;            //  Foam object for generating Initial density
  ofstream *f_Out;              //  External log-file for messages
\end{verbatim}

The most important methods of the {\tt KKee2f} class are the following:
\begin{verbatim}
  void   Initialize(TRandom*, ofstream*, TH1D*);
  void   Generate();
  void   Finalize();
  void   InitParams();
  void   FoamInitA();
  double RhoFoam5(double *Xarg);
  void   Redress(TRandom*, ofstream *, TH1D*);
\end{verbatim}
The {\tt Initialize} method 
transfers all input data into the database {\tt DB} object,
allocates all component objects and initializes them.
The  {\tt InitParams} method helps to set up member data
in the {\tt KKee2f} class during initialization.
The initialization of the {\tt f\_FoamI} 
object in the {\tt Initialize} method memorizes
the {\tt RhoFoam5} distribution. 
It is basically the 
initial state radiation (ISR) function times the resonant Born differential
distribution provided by the {\tt m\_BornDist} object.
Three other methods, 
{\tt MakeGami}, {\tt RhoISRold} and {\tt MapPlus}, are
used in constructing the ISR energy loss function in {\tt RhoFoam5}.
The choice of the final state fermion type 
and the angle $\theta$ of the final fermion is also managed by
the {\tt f\_FoamI} object -- altogether 3 variables.
In the case of the activated beam spread option
two additional variables are also modelled by \foam\
-- that is 5 variables altogether.

The {\tt Generate} method does the most important 
task of generating every single MC event.
It invokes methods of the component objects
{\tt m\_GenISR,  m\_GenFSR } to generate 4-momenta and of the next
objects {\tt m\_QED3, m\_GPS} to calculate the QED matrix element.
The auxiliary class {\tt m\_BVR} provides virtual corrections
and {\tt m\_EWtabs} provides the electroweak formfactors of \dizet\
for the matrix element.
The generated Monte Carlo event is gradually built up
in the {\tt m\_Event} object in the internal format of \kkmcee.
The interface  {\tt m\_HEPMC} 
translates {\tt m\_Event} into a {\tt m\_Hvent} event in the {\tt HEPMC3} format.
The important {\tt m\_TauGen} object manages the calculation
of the MC weight component implementing spin effects in the $\tau$
lepton decays and also transforms tau decay products from the $\tau$
rest frames to the laboratory frame.

The {\tt Finalize} method,
invoked after generation of the MC events, 
prints out the final statistics of the MC run
and the overall normalization (total cross section).
The role {\tt Redress} method in persistency implementation
is described in the following.

\subsection{Persistency mechanism}
Many MC event generators offer the possibility to record
the complete state of the generator in disk files, 
allowing event generation to resume seamlessly after a break in the production process.
For instance, such an option was available in the classic MC event generator 
{\tt BHLUMI}~\cite{Jadach:1996is}.
In \kkmcee, it is possible to write
to disk the entire MC event genrator object,
that is, an object of the {\tt KKee2f} class  
and all its component objects of the auxiliary classes, 
using the persistency mechanism of the CERN ROOT library~\cite{Brun:1997pa}.
It helps that the C++ version of \foam, 
which plays now central role in \kkmcee,
was always compatible  with the ROOT persistency mechanism.

The persistency mechanism is very useful not only for resuming
MC event production after a stop, but also for debugging 
rare problematic MC events, re-using component methods of the MC generator
during later event analysis, and for many other purposes.

Generally, writing objects of C++ classes to disk files is quite nontrivial
because of the use of pointers. 
Our MC event generator has a lot of component objects related/connected 
through pointers.
ROOT does most of the job of the reading C++ objects from the disk,
but the system of pointers between object components in \kkmcee\ 
is too complicated to be restored by ROOT alone.
For this reason, an additional method {\tt KKee2f::Redress} 
is provided to handle the job of properly restoring and correcting
the internal pointer network
between all component objects of the \kkmcee\ event generator.

In particular, there are three special objects residing outside 
the MC generator object of the {\tt KKee2f} class: 
(i) the central random number generator,
(ii) a special histogram holding the overall normalization, 
(iii) a text log file on the disk.
Pointers to them are provided from the main user program
through arguments of the {\tt KKee2f::Initialize}
and {\tt KKee2f::Redress} methods and are distributed among
all component objects allocated in the MC generator object of the {\tt KKee2f} class.

More precisely, the  {\tt KKee2f::Redress} method distributes pointers 
to all component objects 
allocated in the {\tt KKee2f::Initialize} method 
(listed in the previous section) among component objects
according to their functionality.
After restoring the MC generator object of the {\tt KKee2f} class,
it is necessary to use the {\tt KKee2f::Redress} method to correctly reconstruct the original network of pointers.
The main class of \foam\ has its own {\tt Redress} method 
which is invoked by {\tt KKee2f::Redress}.

It is necessary to use {\tt KKee2f::Redress} because,
for instance, the standard ROOT action would be to create copies
of the random number generator within several component objects 
restored from the disk,
while \kkmcee\ uses a single central random number generator 
serving all components of the generator, including \tauola.
The same problem would occur for other interconnected component objects
of the generator allocated originally in {\tt KKee2f::Initialize}
and later restored from the disk.

\subsection{Multiphoton generator classes {\tt KKarlud} and {\tt KKarfin} }
Once the total energy of the 
initial state radiation (ISR) photons is provided by the {\tt f\_FoamI} object,
then the {\tt m\_GenISR} object of the {\tt KKarlud } class generates
the Poissonian multiplicity of ISR photons and next their four-momenta.
The energy-momentum conservation is adjusted by means of rescaling photon
four-momenta, taking into account the respective Jacobian factor, following the method outlined
in Ref.~\cite{Jadach:1988gb}. 
This method is quite efficient except for the corner of the phase space where
two photons are very hard, close to the phase space limit.
Four momenta of all photons are recorded into the {\tt m\_Event}
object of the {\tt KKevent} class.
In the case when 
final state radiation (FSR) is suppressed 
by the switch in the input or in the case of the
neutrino final state, {\tt m\_GenISR} generates also final state fermion momenta,
using $\theta_f$ generated earlier by the {\tt f\_FoamI} object.

The four-momenta of the FSR photons are generated next 
in the {\tt m\_GenFSR} object of the {\tt KKarfin } class.
This is done in the rest frame of the outgoing fermion pair.
FSR photons are then transformed to the laboratory frame and recorded
in the object {\tt m\_Event} of the {\tt KKevent} class.
This class also provides several methods which help to perform Lorentz
transformations on the ISR/FSR photons and fermions.
All four-momenta in {\tt m\_Event} are encoded
using the {\tt TLorentzVector} class of {\tt ROOT} and are handled using
methods of the {\tt KKevent} class.
Let us note that all of the functionality of the {\tt TLorentzVector} class
is the same%
\footnote{This is not surprising as both of them are derived from the
{\tt VECTOR} library of {\tt CERNLIB}.}
as that of the {\tt KinLib} class in the F77 version of \kkmcee.
 
 The {\tt m\_GenISR} and {\tt m\_GenFSR} objects generate ISR and FSR
 multiphoton distributions according to simplified distributions
 which are next corrected by the MC weight according to the matrix
 elements of the {\tt KKqed3} and {\tt KKceex} classes.
 In particular, the initial-final state interference (IFI) is not yet included at this stage.

\subsection{Input data handling with the {\tt KKdbase} class}
The input data in \kkmcee\ are still organized
in the form of one long vector {\tt xpar}. Default values are read from
the {\tt SRCee/KKMCeee\_defaults} file, 
after which the user may overwrite any entry as desired, see Sect.~3.2.
During the execution of {\tt KKee2f::Initialize}, 
these data are transferred to the {\tt DB} object
of the {\tt KKdbase} class. The pointer to {\tt DB} object is distributed
in {\tt KKee2f::Initialize} to all component objects of the generator object.
The member data of the {\tt DB} object do not have the usual {\tt m\_} prefix
in order to facilitate their use in the code.
For instance, the mass of the $Z$ boson can be accessed simply as {\tt DB->MZ}
everywhere in the code.
We have adopted the rule that all member data in the {\tt DB} object are static,
that is they are the same as in the input and do not change during the entire
life cycle of the MC generator object.
This policy is implemented not using C++ language {\tt const} declaration
but as the programmer convention.

Practically all of the description of the input data in Ref.~\cite{Jadach:1999vf}
remains valid and is not repeated here.

\subsection{Matrix element classes {\tt KKqed3, KKceex, KKbvir} and {\tt KKborn} }
The EEX matrix element squared is calculated 
by the {\tt m\_QED3} object of the {\tt KKqed3} class
and CEEX spin amplitudes are evaluated
by the {\tt m\_GPS} object of the {\tt KKceex} class,
using MC event as recorded in the {\tt KKeven} class object.
They provide in the present version of \kkmcee\
exactly the same distributions and/or spin amplitudes as in the F77 version.
Both of them use the EW formfactors of \dizet\ provided by 
the {\tt m\_EWtabs} object of the {\tt KKdizet} class.
This object does not invoke directly \dizet\ but reads
lookup tables from a disk file prepared by the separate program
{\tt MainTab}, which is using the \dizet\ library residing in
a separate directory {\tt dizet}, see Sect.~\ref{sect:dizet}. 
That is why in the calculation of EW formfactors the \dizet\ library can be easily replaced 
with a different library or \dizet\ library version. 

As explained in Ref.~\cite{Jadach:2000ir}, 
the EEX matrix element does not include IFI corrections, 
but provides the complete \order{\alpha^3L_e^3} corrections.
On the other hand CEEX spin amplitudes feature IFI corrections
and also include \order{\alpha^2L_e} corrections.
The calculations of the CEEX spin amplitudes are costly in the CPU time,
partly because the summation over all partitions of photons between initial
and final state emitters has to be performed.
The calculation of the CEEX matrix element requires a lot of
complex calculations on the Weyl spinors.
In order to facilitate these calculations an auxiliary class {\tt KKpart}
was introduced. Its objects combine the particle four-momentum
with the $C$-number saying whether it is a particle or an antiparticle
and with the mass of the particle which enters into a spinor.
Virtual corrections used in both types of matrix elements are
provided by the {\tt m\_BVR} object of the {\tt KKbvir} class.

After the CEEX matrix element is imposed by the MC weight,
then due to quantum mechanical nature of the QED
the distinction between ISR and FSR photons disappears
and the four-momenta of ISR and FSR photons are merged in a single list.

\subsection{The interface to electroweak libraries and \dizet\ electroweak library}
\label{sect:dizet}
The definition of electroweak non-QED one loop effects 
and the corresponding electroweak form-factors was 
the result of a massive LEP-era effort~\cite{ALEPH:2005ab}.
The one-loop level of the LEP-era
was later on shown to be sufficient for the Tevatron/LHC purposes
\cite{Richter-Was:2020jlt}.
Three variants of the \dizet\ used in \kkmcee\ are characterized
in  Ref.~\cite{Arbuzov:2020coe}.
Note that the refinement of the loop corrections necessary for low energies 
is absent in the present version of \kkmcee.
In particular, the results of Ref.~\cite{Banerjee:2007is} 
for hadronic vacuum polarization are not incorporated.

In the present version, we include three versions 
of the \dizet~\cite{Bardin:1989tq} EW library in three separate directories:
{\tt dizet-6.21}, {\tt dizet-6.42} and {\tt dizet-6.45}.
For a more detailed description of these three versions see
Ref.~\cite{Arbuzov:2020coe}.
Before any use of \kkmcee\ one has to manually link
the preferred directory to the {\tt dizet} directory,
for instance: {\tt link -s dizet-6.45 dizet}.

The \kkmcee\ event generator communicates with the \dizet\ F77 library
through the text file {\tt DIZET-table1} created by the {\tt TabmainC} program
located in a separate {\tt DZface} directory.
The source of {\tt TabmainC} consists of only one small C++ program {\tt TabMain.cxx}
and one small F77 program {\tt hhDizet.f}, which 
both reside in the {\tt DZface} directory.

Before starting the \kkmcee\ event generator, the {\tt DIZET-table1} file
must be created in the local directory by invoking {\tt TabmainC}
using the same local input file as \kkmcee.
The {\tt DIZET-table1} file could be reused, 
but we recommend recreating it before any new MC run 
(in the user examples of the next section, {\tt Makefile} recreates this
file automatically).
The above organization is a slightly simplified version of the interface to \dizet\
in the classic \kkmc~\cite{Jadach:2000ir}
and most likely will stay quite similar in the case of any new EW library in the future.
In particular, two-loop EW corrections (once available)
will cost a lot of CPU time to calculate and will have to be stored on disk
in the form of predefined look-up tables.

\subsection{The interface to $\tau$ decay library \tauola}
Let us describe the present status of 
the entire interface to the $\tau$ decay library.
It is quite complicated and exploits several components of \kkmcee.
The entire process of generating $\tau$ decays
and implementing spin effects in the decays is managed
by the following sequence of the code in the {\tt Generate()}
method of the main generator object of the {\tt KKee2f} class:
\begin{verbatim}
  m_GPS->TralorPrepare(1);  // prepare transformations tau frame -> LAB
  m_GPS->TralorPrepare(2);  // accounting for tau spin quantization axes
  m_TauGen->DecayInRest();  // tau decays (f77 Tauola) in tau rest frames
  m_TauGen->ImprintSpin();  // implementing spin effects
  m_TauGen->TransExport();  // transform decays to LAB, collect tau decays
  m_HEPMC->tauolaToHEPMC3();// append  m_Hvent with tau decay products
  m_TauGen->RunPhotosPP();  // Run Photos for non leptonic tau dacays
\end{verbatim}
When implementating spin polarization effects, it is critical to know
precisely the directions of all three axes of each $\tau$ 
rest frame used to quantize the spin of the $\tau$.
The prescription for finding these axes for the CEEX spin amplitudes
calculated in the {\tt m\_GPS} object was given in Ref.~\cite{Jadach:1998wp}.
Here, the {\tt m\_GPS->TralorPrepare} method finds these frames and records the
parameters of the Lorentz transformation from 
both $\tau$ rest frames with the proper quantization axes 
to the laboratory frame for later use.

In the next step, {\tt m\_TauGen->DecayInRest()} performs decays of both $\tau$'s
in their rest frames using the {\tt DEKAY} F77 subroutine of \tauola.
The {\tt DEKAY} routine returns polarimeter vectors dependent on the decay momenta,
which are used in {\tt m\_TauGen->ImprintSpin()} to calculate the MC weight
implementing spin polarization effects, 
including spin correlations between the two decays.
The spin weight is calculated using the {\tt m\_GPS->MakeRho2} method
of the {\tt KKceex} class, inside the {\tt TauPair::ImprintSpin()} method.
The additional randomisation of the decay products using random
Euler rotation, similar as in the original \kkmc\ program,
is also performed by the {\tt TauPair::RandRotor()} method.

Next, the {\tt m\_TauGen->TransExport()} function is invoked.
This part is a complicated mixture of the actions of F77 and C++ programs.
The {\tt DEKAY} fortran subroutine of \tauola\ transforms
decay products of both $\tau$'s from the $\tau$ rest frames to the laboratory frame
using the {\tt TRALOR4} subroutine.
However, the {\tt DEKAY} is in fact using 
the transformation of the C++ function {\tt m\_TauGen->Tralo4(...)}
through a wrapper defined in the {\tt SRCee/Globux.h} interface:
\begin{verbatim}
  void tralo4_(int *KTO, float P[], float Q[], float *AM){
      g_KKeeGen->m_TauGen->Tralo4(*KTO, P, Q, *AM); }
\end{verbatim}
The {\tt TRALOR4} wrapper subroutine is called in many places in \tauola.
In the same places of the \tauola\ code, 
another {\tt FILHEP3} wrapper subroutine is executed, 
which in fact is a wrapper to the C++ function {\tt  HepFace::FillHep3}.
Its role is to collect a list of all decay products,
which is used in the next call {\tt m\_HEPMC->tauolaToHEPMC3()}
 to append the {\tt HEPMC3} event record.
(The {\tt m\_HEPMC} event was already filled 
earlier with the fermions and photons
of the fermion pair production process.)

Finally \photos\ is invoked in {\tt m\_TauGen->RunPhotosPP()}.
It processes the {\tt m\_HEPMC} event, adding photon emission to all
charged $\tau$ decay products in all decay modes except leptonic
$\tau$ decays, since \tauola\ already adds photons for these.
The decision of whether or not to invoke \photos\ for a given final state particle is made
using a dedicated ``filter'' in the {\tt TauPair::RunPhotosPP()} method rather than using the steering parameters of \photos.

The latest \photos\ version
has also the interesting option of generating light
lepton pairs independently of the photon emission, 
not only for $\tau$ decay products but also for any final fermions.
For the moment this option is not active.
Using it would require modifying the ``filter'' 
in the {\tt TauPair::RunPhotosPP} method.

\subsection{\kkeefoam\ semi-analytical tool}
The \kkeefoam\ program provides predictions for the total cross section
and charge asymmetry using a combination of analytical
 Monte Carlo calculations.
The partial analytical integration over the multiphoton phase space,
including the IFI component, is described in Ref.~\cite{Jadach:2018lwm}.
The numerical integration over the remaining phase space is done using \foam.
The source code of \kkeefoam\ is in the {\tt SRC/KKeeFoam.cxx} file.

The \kkeefoam\ code was in C++ from the beginning.
Its full algebraic content is described in Ref.~\cite{Jadach:2018lwm}.
Here we add only some details on its components and MC algorithm.
Its use will be described in the next section.

Although \kkeefoam\ is not a true MC event generator providing four-momenta,
its use is similar, because it provides MC events with
the effective mass and azimuthal angle of the final fermion pair.
Internally, it also generates longitudinal momenta of the ISR and FSR photons.
Photon transverse momenta are integrated analytically.
When IFI is switched on, \kkeefoam\ integrates also
numerically over two additional convolution variables, 
see Ref.~\cite{Jadach:2018lwm}. 
In this case, the MC weights of \kkeefoam\ events are non-positive.
Since the integrand of \foam\ is significantly different in the cases
of IFI on/off, \kkeefoam\ embeds two different \foam\ objects.
The single MC event of  \kkeefoam\ in fact consists of two
MC events provided by two \foam\ objects, one with IFI and another one without IFI.
As in the case of \kkmcee, the \kkeefoam\ class inherits
from the classes of the {\tt MCdevelop} library~\cite{Slawinska:2010jn}.

\section{Use of the program}
In the \kkmcee\ distribution directory, we include two types of
user examples, a very simple one which generates MC events only
and a few more advanced examples which reproduce benchmarks
and/or can be used as a template for a sophisticated analysis
of the \kkmcee\ results.
We begin with instructions on how to compile and link
all libraries and executables of the program.

\subsection{Building the program}
It is quite easy to build
all of the shared libraries and executables of the project using automake/autotools:
\begin{verbatim}
     ln -s dizet-6.45 dizet
     autoreconf -i --force
     ./configure CXXFLAGS="-std=c++11 -g -O2" 
        --with-hepmc=/opt/hepmc3-install
        --with-photos=/opt/PHOTOS-install
     make
\end{verbatim}
Note that ”-std=c++11” option is a special case needed only
if one uses an old C++ compiler, 
which defaults to ”-std=c++98” 
(e.g., g++ 4.8.5 on CentOS 7), together with an old-fashioned
ROOT 5 (which, by default, is also built using c++98 compatible mode).
Otherwise, setting the C++ standard manually is not necessary --
it will be automatically ”inherited” from ROOT.

It is necessary to link manually one of the three versions of the \dizet\ directory 
in preparation for the build, for instance by executing the command
{\tt 'ln -s dizet-6.45 dizet'}  
in the main directory of the distribution.
Switching to another version of \dizet\ requires re-linking the \dizet\ directory
and rebuilding the project from the scratch.
Locations of the external \hepmc3 and \photospp\ libraries must be provided
explicitly in the {\tt configure} parameters.

In some systems, the automake system may require appending
environmental variables, for example one may need to add 
the path to \photospp \ library in bash:\\
{\tt export LD\_LIBRARY\_PATH=\$LD\_LIBRARY\_PATH:PHOTOS-install-path/lib/}.

During construction and testing the present version of the program we have
used version 6.x of {\tt ROOT}, 
version 3.64 of \photos\ from\\
\centerline{\href{https://gitlab.cern.ch/photospp/photospp}%
{\tt https://gitlab.cern.ch/photospp/photospp}}
and version 3.2.2 of \hepmc3 from\\
\centerline{\href{https://gitlab.cern.ch/hepmc/HepMC3}%
{\tt https://gitlab.cern.ch/hepmc/HepMC3}}

Installing auxiliary libraries%
\footnote{Note that \hepmc3, \photospp\ are published and documented 
in Computer Phys. Commun. \cite{Buckley:2019xhk,Davidson:2010ew}.}
\hepmc3, \photospp\ and  {\tt ROOT} in the operating system of the user, 
or locally in the user account, can be a non-trivial task.
An interesting alternative is to import from the CERN web site
a ready-to-go virtual machine system {\tt CernVM}
with all the above auxiliary libraries already preinstalled.
The virtual machine {\tt CernVM} provides a
{\em sophisticated installation framework} for \kkmcee.

The detailed instruction on how to install {\tt CernVM} virtual machine
with the preinstalled CENTOS 7 operating system 
can be found in section 1.1.3.2 on the following webpage\\
\centerline{\href{https://hep-fcc.github.io/fcc-tutorials/software-basics/README.html}%
{\tt  https://hep-fcc.github.io/fcc-tutorials/software-basics/README.html}}.
Under {\tt CernVM} the following simple command sequence builds the project:
\begin{verbatim}
     ln -s dizet-6.45 dizet
     autoreconf -i --force
     ./configure
     make
\end{verbatim}

\subsection{Running a simple example }
The source code of a very 
simple example user program 
{\tt ProdRun/MainMini.cxx},
which generates a series of the MC events using \kkmcee, appears as follows:
\begin{verbatim}
#include "TRandom1.h"
#include "TH1.h"
#include "KKee2f.h"
int main(){
ofstream   OutFile("pro.output",ios::out);  // Logfile output
TRandom1 *RN_gen = new TRandom1();// Central random numb. gen.
long iniseed = 54217137; RN_gen->SetSeed(iniseed);
KKee2f *KKMCgen = new KKee2f("MCgen"); // MC generator object
int nb = 10000;
TH1D *h_NORMA= new TH1D("KKMCgen_NORMA","Normaliz. histo",nb,0,nb);
KKMCgen->Initialize( RN_gen, &OutFile, h_NORMA);
int NevGen =100;
cout<<"MainMini: ********************************** "<<endl;
cout<<"MainMini: type in no. of MC events: (100?) ";
cin>>NevGen; cout<<" requested "<< NevGen <<" events"<<endl;
// Loop over MC events
for(int iev=1; iev<=NevGen; iev++) {
   if( (iev/20000)*20000 == iev) cout<<" iev="<<iev<<endl;
   KKMCgen->Generate();
}// for iev
KKMCgen->Finalize(); // final printout
cout << "  |  TestMini Ended   | "<<endl<<flush;
return 0;
}// main
\end{verbatim}
The above code does not require much explanation.
One can see, for instance, that the central random number generator
object {\tt RNgen} provided by the {\tt ROOT} resides outside the MC generator
and the {\tt KKMCgen} object gets the pointer of the {\tt RNgen}.
The other pointer needed by {\tt KKMCgen} is to the 1-dimensional
histogram {\tt h\_NORMA}, which holds information on the overall normalization
of the MC generator, to be used in the analysis of the generated MC events
in more sophisticated scenarios.

In order to compile and run {\tt MainMini}, the user may execute:
\begin{verbatim}
    cd ProdRun/workMini
    cp workMini.input_105GeV_Leptons workMini.input
    make start
\end{verbatim}
The user will be prompted to provide the number of events to be generated
and the execution will start.
The output file {\tt pro.output} contains the MC result for the total cross section
and the multitude of statistics on the MC generation process.

The {\tt 'make start'} command creates the {\tt DIZET-table1}
file locally by executing the command sequence
{\tt '../../DZface/TabMainC; ../MainMini'}.
Both the {\tt TabMainC} executable and the {\tt KKMCgen->Initialize()} function
read the same input data {\tt ./pro.input},
which is copied from {\tt workMini.input} by {\tt workMini/Makefile}.
Finally, the executable {\tt MainMini} must have access to
the default input file {\tt ./KKMCee\_defaults}.
However, the necessary soft link to {\tt ../../SRCee/KKMCee\_defaults} 
is automatically created by the {\tt workMini/Makefile}.

Input parameters of \kkmcee\ are organized the same way as in the original F77 code
and most of them have the same meaning.
Default values of all parameters which define physics and control running of the program
are kept and defined outside the code in the {\tt SRCee/KKMCee\_defaults} file
of the source distribution,
which is read in the initialisation phase of the execution.
The other small input data file {\tt ./pro.input} provided
by the user is read shortly afterwards and contains
only a small subset of the input parameters redefined by the user, for instance the total
center of mass energy or the list of active final state channels.
Most of the input parameters in the {\tt CRCee/KKMCee\_defaults} are the same as in Tables 2-5
in Ref.~\cite{Jadach:2000ir} and we are not reproducing them in the present article.
The best way to learn about the exact meaning of all input parameters 
of the present version of the programs is to
look into the comments in the {\tt SRCee/KKMCee\_defaults} file.

There is also an additional
simple example user program {\tt ProdRun/MainPers.cxx}
which demonstrates the use of persistency mechanism.
After running  {\tt MainMini} one may continue MC generation
using {\tt MainPers} executable as follows:
\begin{verbatim}
    cd ProdRun/workMini
    make start2
\end{verbatim}
{\tt MainMini} will read MC generator object from the disk file
and continue MC event generation, as if there was no execution break.

In case user is interested to use \hepmc\ event record,
he may access it from the main user program through public pointer
{\tt KKMCgen->m\_Hvent}.

\subsection{Advanced user program examples}
The directories {\tt ProdRun} and {\tt ProdDigest} contain
more sophisticated examples of the user programs.
{\tt ProdRun} provides an example of generating MC events using the {\tt MCdevelop} toolbox,
while {\tt ProdDigest} contains programs for analysing these MC events.
All programs in {\tt ProdRun} implement and exploit
the {\em persistency} mechanism provided by {\tt ROOT}, which allows
C++ objects to be written into and read from disk files.

In {\tt ProdRun}, we have a standardized universal 
{\tt MainMCdev.cxx} main program which refers only to baseline virtual
classes of the {\tt MCdevelop} library such as {\tt TMCgen} and {\tt TRobol}.
It also manages input and output files, in particular the {\tt ROOT} files 
storing histograms and MC generator objects.

The actual analysis programs which run the
MC generators and accumulate histograms
are  {\tt TRobolKKMC.cxx} and {\tt TRobolFoam.cxx}, which
inherit from the {\tt TRobol} class of {\tt MCdevelop},
The former one uses the \kkmcee\ main MC generator and the latter
one employs the auxiliary \kkeefoam\ tool.
{\tt ProdRun} contains three subdirectories
which serve as local directories with input and output files during MC runs,
{\tt ProdRun/work1}, {\tt ProdRun/NU} using
{\tt TRobolKKMC} (for \kkmcee)
and {\tt ProdRun/Foam} using {\tt TRobolFoam} (for \kkeefoam).
Note that an example of the use of {\tt KKMCgen->m\_Hvent} pointer
to MC event in the \hepmc\ format is present in {\tt TRobolKKMC.cxx}.

Running \kkmcee\ in the folder {\tt ProdRun/work1} is very simple:
\begin{verbatim}
      cd ProdRun/work1
      cp work1_189GeV.input work1.input
      make start
\end{verbatim}
The commands for starting MC runs in {\tt ProdRun/workNU} and {\tt ProdRun/workFoam}
are the same except for different input data names.
The simple {\tt 'make start'} command encompasses the following sequence of steps:
\begin{verbatim}
make start -n
(ln -s ../../SRCee/KKMCee_defaults ./)
(cd ../..; make)
cp ./work1.input ./pro.input
(../../DZface/TabMainC;)
(cd ../..; make)
rm -f ./pro.output ./histo.root ./mcgen.root
cp ./work1.input ./pro.input
/usr/bin/root  -b -q -l ./Start.C
../MainKKMC &
\end{verbatim}
First, a soft link to the default input file {\tt KKMCee\_defaults} is created
(in case it is not there). 
Then, the project build is checked for completeness.
After some cleanups, the user input data file {\tt ./pro.input } is prepared.
The execution and the role of the {\tt Start.C} script is explained below.
Finally, the main program {\tt ../MainKKMC} is initiated to generate the MC event series.
The program stops when the number of requested MC events 
defined in {\tt Start.C} is reached. 
Alternatively, issuing the command {\tt make stop} will terminate 
a run cleanly any time while it is in progress.

The script {\tt Start.C} plays an important role  
in the {\tt MCdevelop} scheme.
It creates objects of the MC event generator class \kkmcee\
and of the {\tt TRobolKKMC} class and writes them into the disk file 
{\tt mcgen.root}.
{\tt Start.C} also creates the central random number generator
object {\tt RNgen}.
All these objects are written into {\tt mcgen.root} for subsequent use.
The auxiliary ``sempahore'' object of the {\tt TSemaf} class is
to be used for starting and stopping the MC generator during the MC run.
It is written into the {\tt semaf.root} file.
In addition, {\tt Start.C} also creates the {\tt histo.root} file 
for storing {\tt ROOT} histograms.

The {\tt MainKKMC} executable
reads all objects created by {\tt Start.C} from disk files,
processes them further, and runs the main loop of the production of MC events.
The {\tt MainKKMC} executable
is a part of the MC production and analysis chain consisting
of {\tt TRobolKKMC.cxx} and the analysis programs
{\tt Plot1.cxx}, {\tt PlotNU.cxx}, {\tt PlotBES.cxx} , {\tt PlotTau.cxx} 
and {\tt TabIIIPRD.cxx}
residing in the {\tt ProdDigest} directory.
During compilation, {\tt TRobolKKMC.o} is linked with many other shared libraries
to make the {\tt libProdKK.la} library.
{\tt MainKKMC} is obtained from linking {\tt MainMCdev.o} and {\tt libProdKK.la}.

The other main executable {\tt MainFoam}, which is run in the {\tt ProdRun/workFoam},
is obtained from {\tt MainMCdev.o} and the {\tt libProdFoam.la} shared library.
This library encompasses {\tt TRobolFoam.cxx}.
Results from {\tt MainFoam} in {\tt ProdRun/workFoam} are used in the analysis
programs to compare with the results from the main \kkmcee.
Note that we do not provide separately any simplified main program 
for running \kkeefoam\ similar to {\tt MainMini}.
The procedure for running {\tt MainFoam} in {\tt ProdRun/workFoam} is as follows:
\begin{verbatim}
      cd ProdRun/workFoam
      cp workFoam_189GeV.input workFoam.input
      make start
\end{verbatim}

\begin{figure}[!h]
\centering
\includegraphics[width=0.80\textwidth]{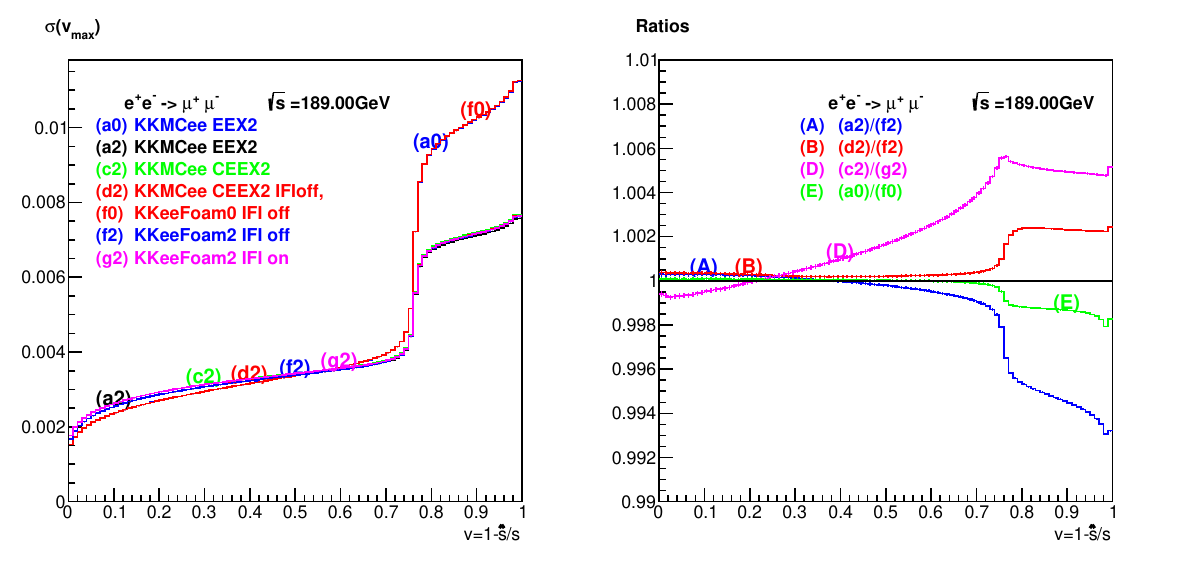}
\includegraphics[width=0.80\textwidth]{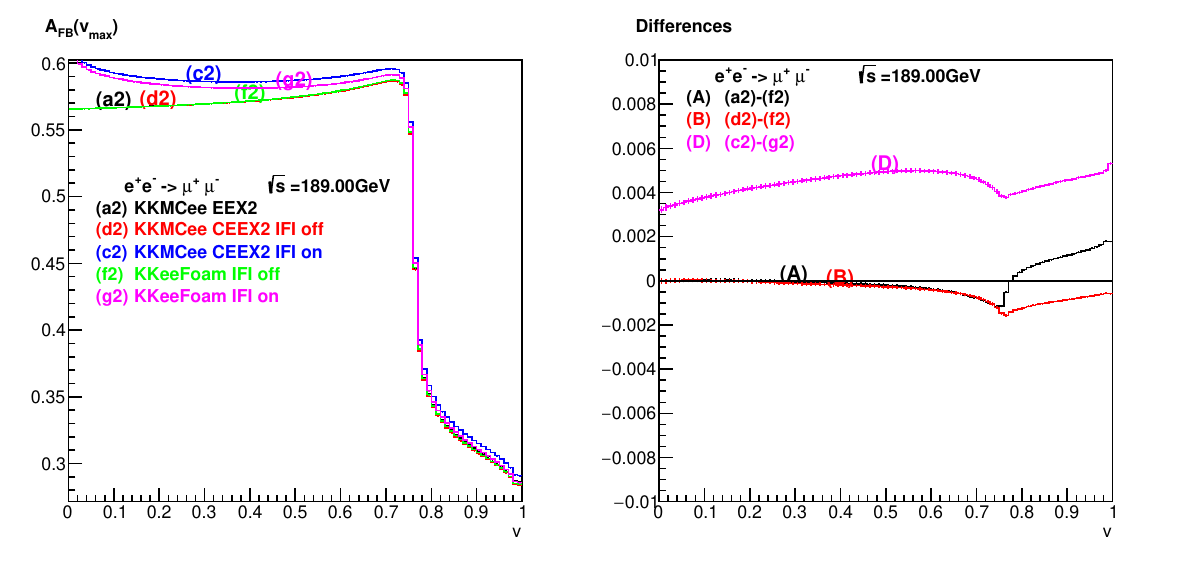}
\caption{
Sample results comparing various types of QED matrix element for {\tt KKMCee} 
and {\tt KKMCeeFoam} generated via the {\tt ProdDigest} analysis programs.
}
\label{fig:Plot1}
\end{figure}
Let us now describe briefly the analysis programs  
in the {\tt ProdDigest} directory.
Their purpose is to analyse MC results, producing many plots and tables 
-- some of them reproduce important basic benchmarks of \kkmcee.
For instance, {\tt `cd ProdDigest; make Plot1'} produces
plots of $\sigma_{\rm tot}(v_{\max})$ 
shown in Fig.~\ref{fig:Plot1}, which compare results from \kkmcee\ and \kkeefoam\
for various types of QED matrix element, similar to Fig.~22
in Ref.~\cite{Jadach:2000ir}, 
where results of \kkmc\ were compared with those from {\tt KKsem}.

The following source code resides in the {\tt ProdDigest} directory:
{\tt Plot1.cxx} plots comparisons of the  cross sections and charge asymmetry from
\kkmcee\ and \kkeefoam\ for the $\mu$-pair final state,
while {\tt TabIIIPRD.cxx} produces the same output in tabular (PDF) form; 
see Sect.~\ref{sec:benchm}.
In fact {\tt TabIIIPRD.cxx} produces \LaTeX\ file to be processed
by \LaTeX, hence \LaTeX\ should be available in the system.
The program {\tt PlotNU.cxx} performs similar comparisons 
and other analyses for neutrino pair final states.
The {\tt PlotTau.cxx} reproduces benchmarks on spin effects in $\tau$ decays.
Finally, {\tt PlotBES.cxx} tests distributions of the beam energies
when the option of beam energy spread or beamstrahlung is switched on.
More numerical results from these analysis programs will be shown
in Sect.~\ref{sec:benchm}.

\subsection{Running on the farm}
The \kkmcee\ distribution directory includes a set of scripts
and {\tt Makefile}'s for the MC production on a PC-farm, which 
(i) create separate subdirectories for input/output
files for each batch job, 
(ii) send jobs into the execution queue,
(iii) control running jobs and 
(iv) collect {\tt ROOT} output files from many jobs into a single file.
In the F77 version those scripts were written in C shell and now
they are transformed into C++ scripts run by {\tt ROOT}.
This is in order to facilitate read/write access to the semaphore objects,
which are controlling the running MC batch jobs.
All farming scripts reside in the {\tt MCdev/farming} directory.
The example of the command sequence which initiates a series of 24 batch jobs
on the farm looks as follows:
\begin{verbatim}
    cd ProdRun/work1
    make SLfarm24
    make SLsubmitall
    make q-nev
    make q-sem
    make farm-stop
    make combine
\end{verbatim}
The meaning of these commands is self-explanatory: in particular
{\tt `make q-nev'} and {\tt `make q-sem'} are queries on the running jobs
and {\tt `make combine'} is combining all files {\tt work1/farm/*/histo.root}
into a single file {\tt work1/histo.root}.
This can be done even without stopping batch jobs, because
the {\tt MainKKMC} dumps all histograms into a local {\tt histo.root}
file after generating every {\tt ngroup} number of events
(the variable {\tt ngroup} is defined in {\tt Start.C}).
The names of the batch system commands and parameters must be customised
manually in the farming script codes and/or 
in {\tt Makefile.am} files for any particular batch system.

\section{Improvements in the MC algorithm}
The basic algorithm on generating photon four-momenta in \kkmceeFive\ 
is the same as in the classic F77 version.
Nevertheless, some improvements in the algorithm were introduced,
 mostly related to a wider use of \foam.

\subsection{The central use of the {\tt FOAM} tool}
\begin{figure}[!h]
\centering
\includegraphics[width=0.80\textwidth]{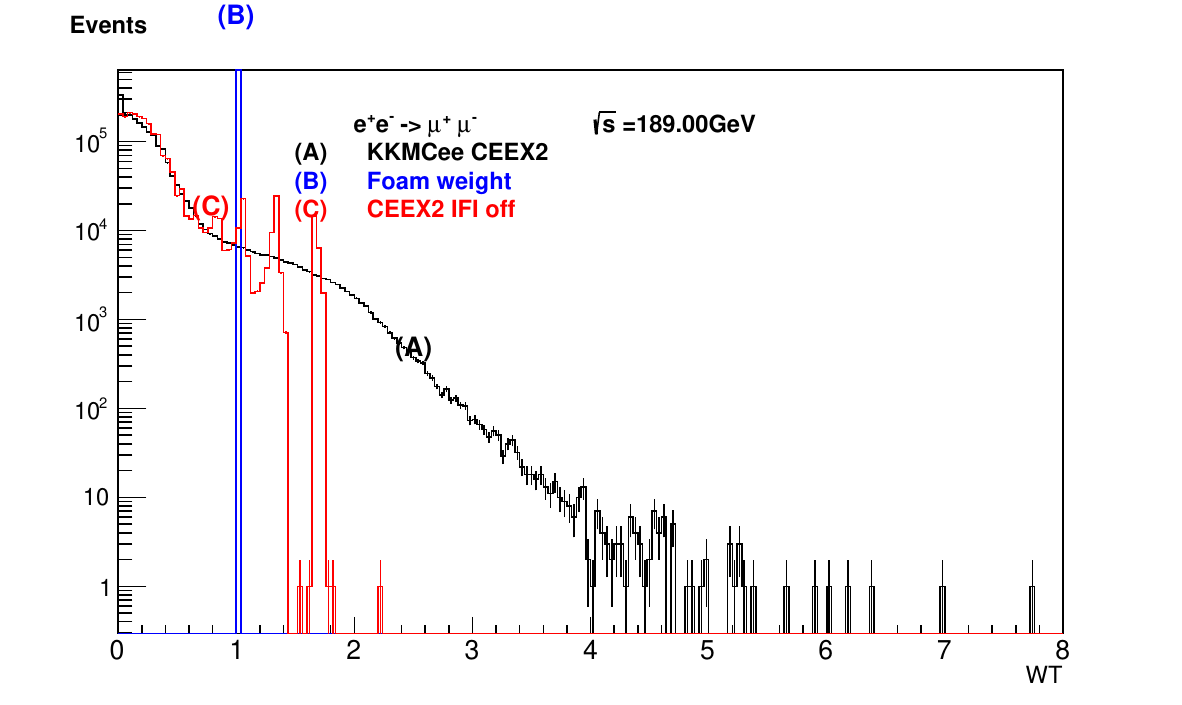}\\
\includegraphics[width=0.80\textwidth]{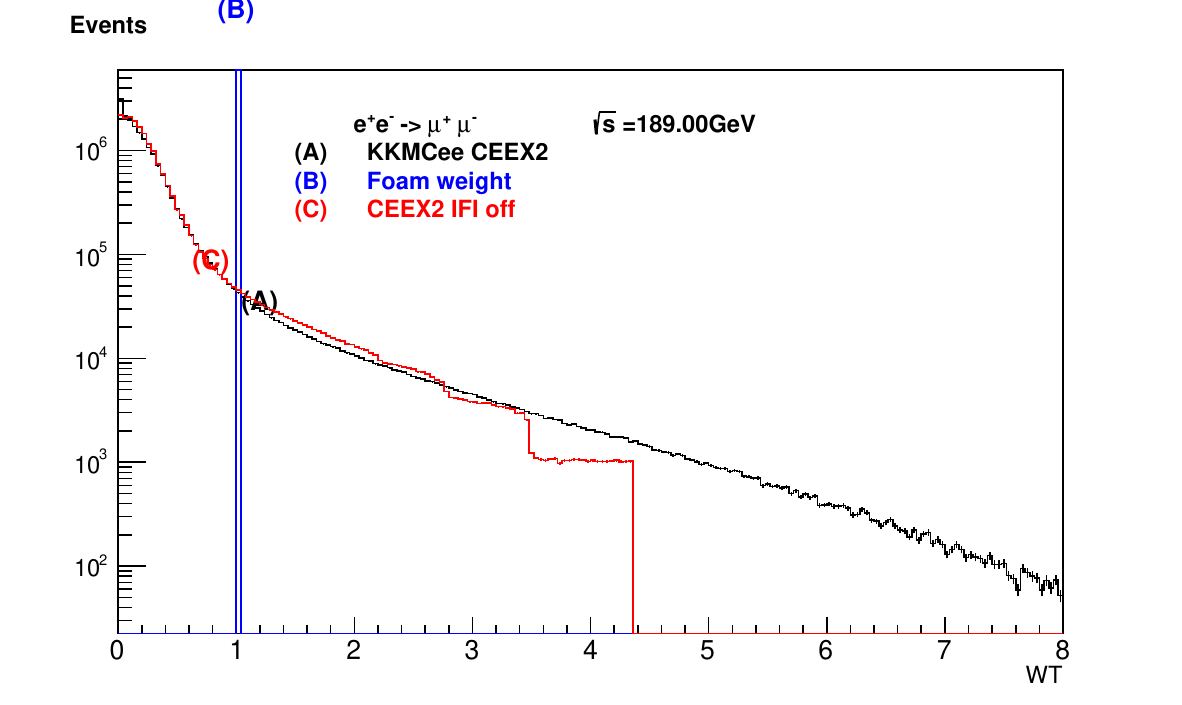}
\caption{
Curve (A) shows the weight distribution for the muon pair channel at $189$~GeV for a
MC sample of 2M variable-weight events using \kkmceeFive\ (upper panel) and Fortran \kkmcee \ (bottom panel). Curve (B) is Foam weight (which is the Dirac delta)
and curve (C) shows the weight distribution with IFI switched off.}
\label{fig:A0_proposal}
\end{figure}

The general-purpose Monte Carlo tool \foam\ of 
Ref.~\cite{Jadach:1999sf} is a self-adapting Monte Carlo tool
able to simulate (integrate over) an arbitrary multi-dimensional distribution
provided by the user function.
\foam\ begins by exploring and memorizing the user distribution in 
a short exploratory MC run, before its actual use in event generation.
During this exploration, it 
creates a grid of rectangular cells, 
denser in the regions where the user distribution is enhanced.
\foam\ has a lot of steering parameters, which in \kkmcee\ are accessible
to the user through input parameters.
However, all steering parameters are set to some well chosen default values
and there is really no need to adjust them any further.
The \foam\ can provide variable weight events or weight one events --
in the \kkmcee\ we have chosen to use it in the weight-one mode,
in order to improve CPU efficiency
(this can be changed using input parameters).
\foam\ also provides the basic overall normalization of the \kkmcee\ weights.
It requires a mapping of the user parameter space into a unit hypercube.
The quality of the grid of \foam\  cells from the initial exploration 
and the quality of the \foam\ weight distribution
can be improved with the help of a mapping
which takes into account peaks in the user function.
In \kkmcee\ the variable mappings done for the \foam\ are quite sophisticated,
although it was not really mandatory to do it, because \foam\ is able 
to automatically find and account for 
the peaks in the user distribution quite efficiently.

The \foam\ was already introduced into the F77 version of KKMC
some time ago in order to generate incoming beam energies 
due to the beamstrahlung effect in high energy lepton linear colliders. In this application,
the F77 version of \foam\ handled three variables, $z_1$ and $z_2$
of the beamstrahlung spectrum and $v$ parametrizing the total energy of the ISR photons.
In the present \kkmceeFive\ program, the C++ version 
of \foam~\cite{Jadach:2002kn}
is used to generate not only the above three variables 
but also the type of the final fermion 
$f=\mu,\tau,\nu_e,\nu_\mu,\nu_\tau,u,d,s,c,b$ 
and the angle $\theta_f$ for the final fermion with respect of the initial beams.
In the F77 version, $\cos\theta_f$ was primarily generated 
with a flat initial distribution,
to be modelled later by the main MC weight.
Shifting $\theta_f$ generation to the \foam\ level saves CPU time
because \foam\ can generate weight-one events%
\footnote{Calculation of the CEEX matrix element takes much more time than \foam.}
more efficiently.
This is because the distribution of the main MC weight has a smaller dispersion
and a nicer tail of the weight distribution --
this is important in the rejection procedure 
turning weighted events into weight one events.
This can be seen in
Figs.~\ref{fig:A0_proposal}, where we compare the weight distributions
for the muon pair production channel in the present C++ version (upper panel) 
and in the classic Fortran \kkmc \ (bottom panel).
The weight distribution in the case of IFI switched off is also shown there.
The weight component of \foam\ is trivially peaked at $W=1$.

Generating $\cos\theta_f$ according to the Born cross section using \foam\
improves the MC efficiency and the shape of the weight distribution
even more in the electron neutrino channel.
This is because above $105$~GeV, the distribution of $\cos\theta_f$ 
starts to feature a quite strong peak at $\theta_f=0$
due to $t$-channel $W$ exchange.
More discussion on all the above questions is presented 
in the Appendix~\ref{apx:mc-weight}.

\subsection{Generation of the beam energy spread}
\begin{figure}[!h]
\centering
\includegraphics[width=0.49\textwidth]{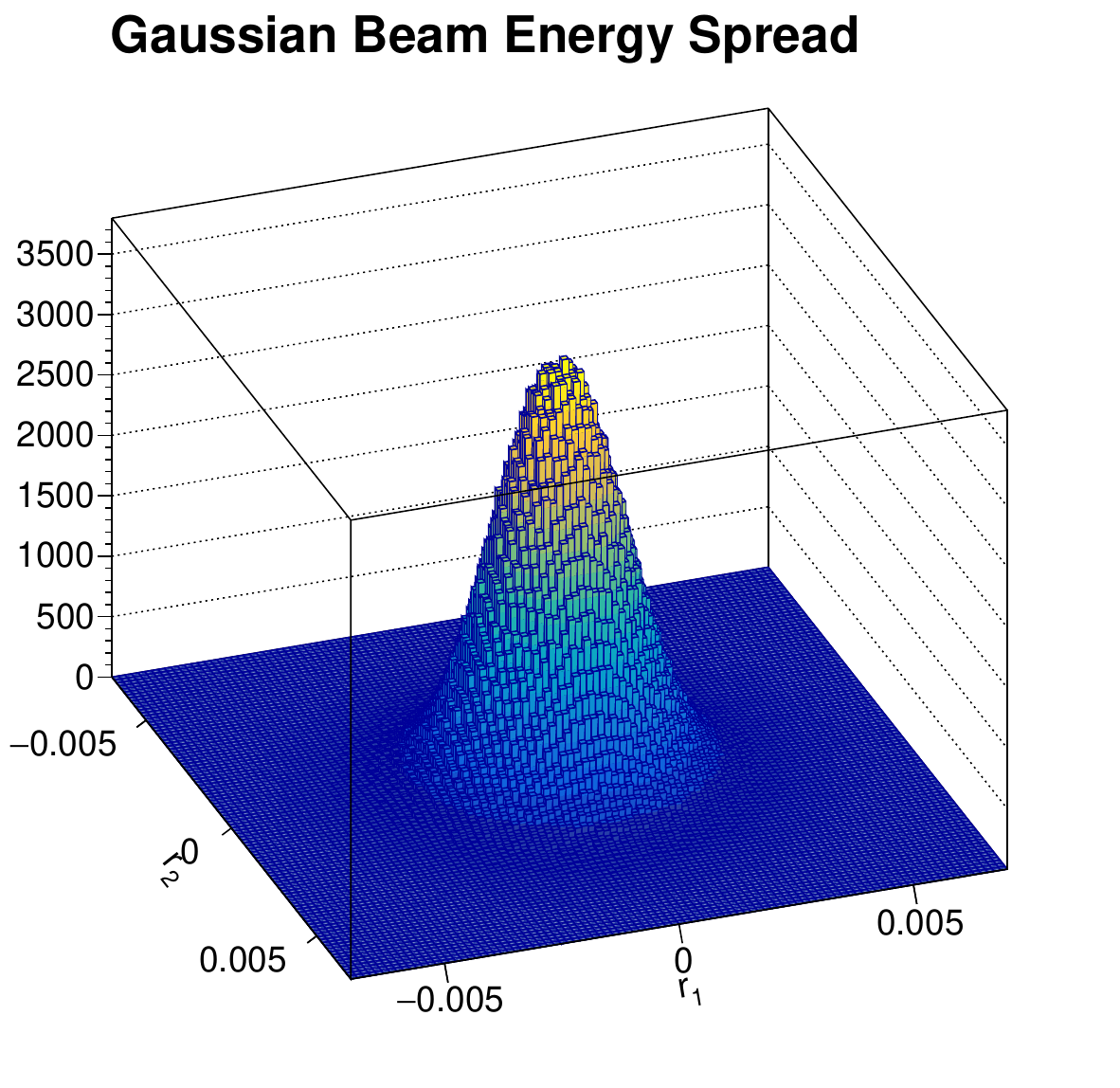}
\includegraphics[width=0.49\textwidth]{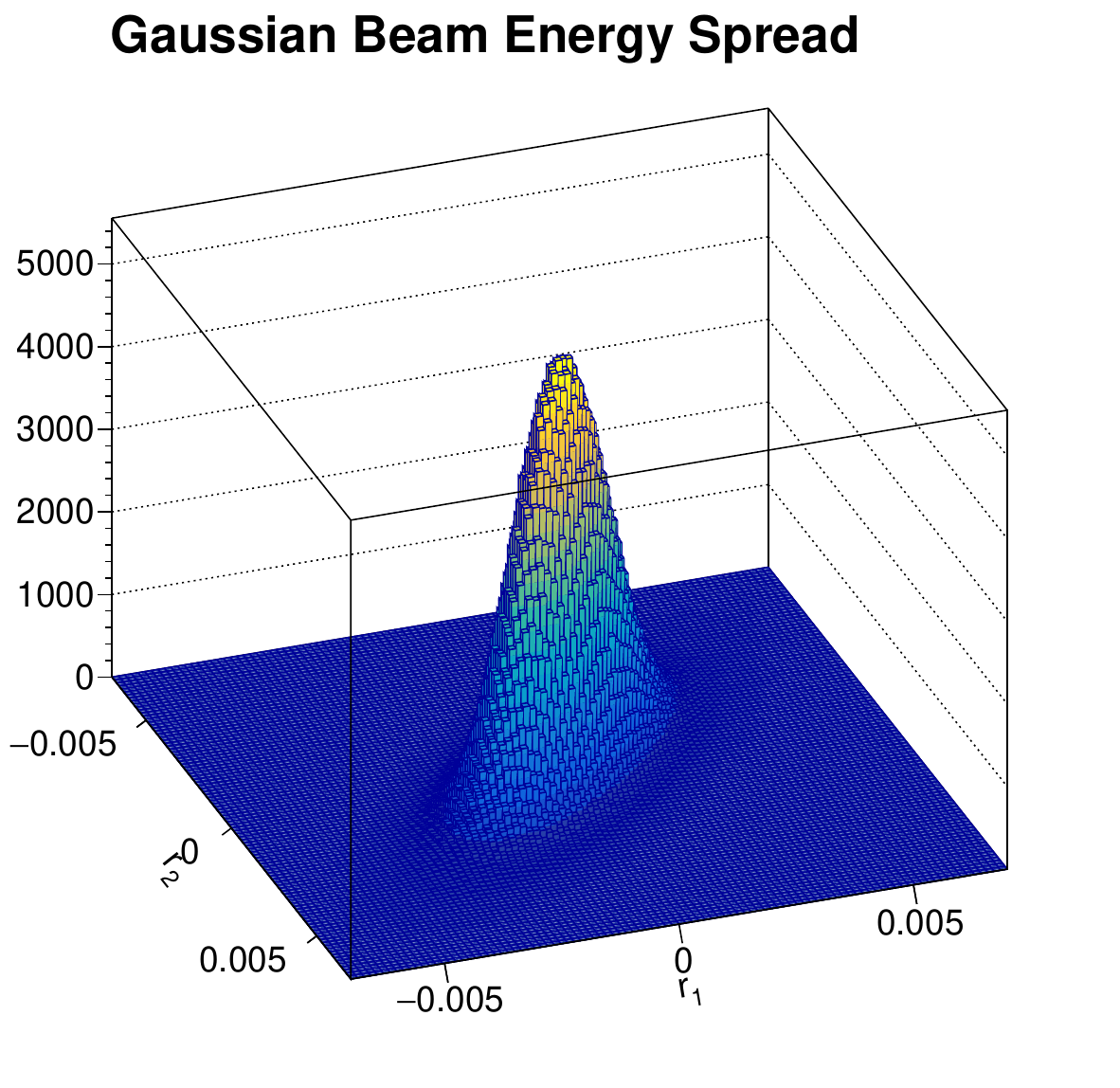}
\caption{
Beam energy spread of two beams
as seen in the variables $r_i=(E_i-\bar{E}_i)/\bar{E}_i$.
}
\label{fig:BES}
\end{figure}
Gaussian beam energy spread (BES) is included optionally
in the present version of \kkmcee.
It is generated using \foam\ according to a strongly
correlated double-Gaussian:%
\footnote{ We thank Patrick Janot for providing us with this distribution.}
\begin{equation}
P(E_1,E_2)= \frac{1}{2\pi\sigma_1\sigma_2\sqrt{1-\rho^2}}
\exp\left\{ \Big( \frac{E_1-\bar{E}_1}{ \bar{E}_1 \sigma_1} \Big)^2 
             +\Big( \frac{E_2-\bar{E}_2}{ \bar{E}_2 \sigma_2} \Big)^2 
-2\rho  \frac{E_1-\bar{E}_1}{ \bar{E}_1 \sigma_1} \; \frac{E_2-\bar{E}_2} { \bar{E}_2 \sigma_2}
\right\}.
\end{equation}
The example default parameters for FCCee are optimistically taken to be
$\sigma_1=\sigma_2= 1.32\times 10^{-3}$ and $\rho= -0.745$.
They can be changed by the user in the input data.

Fig.~\ref{fig:BES} shows the energy distribution of two beams
from a short MC run of \kkmceeFive.
The LHS plot uses the correlation parameter $\rho=0$,
while the RHS plot is for $\rho= -0.745$.
Both plots use the default values of $\sigma_i$.

\subsection{Beamstrahlung}
\begin{figure}[!h]
\centering
\includegraphics[width=0.55\textwidth]{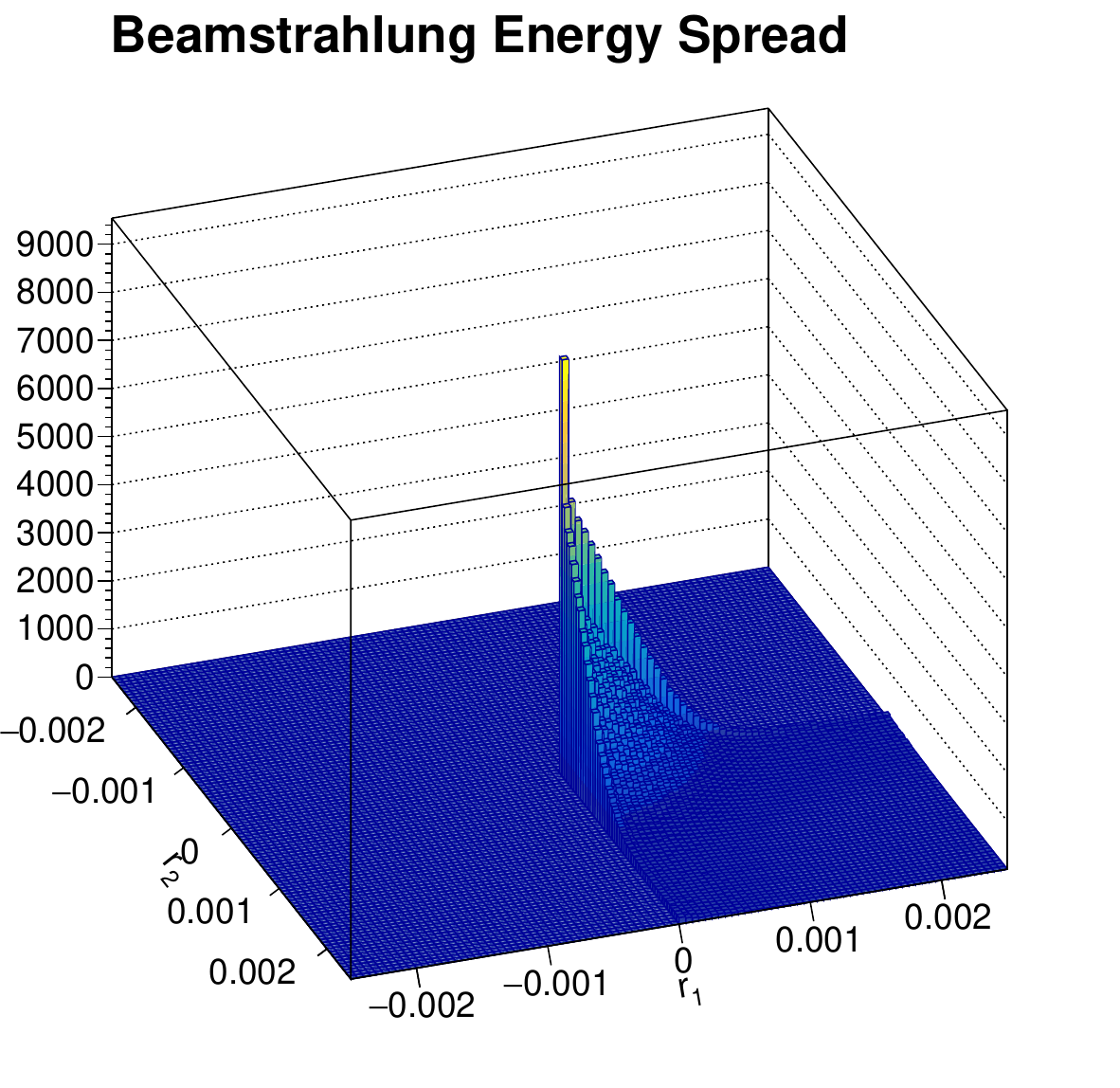}
\caption{
Beam energy spread due to beamstrahlung
according to Eq.~(\ref{eq:BST}), $r_i=1-z_i$.
}
\label{fig:BST}
\end{figure}
Beam energy spread due to beamstrahlung (BST),
which can be quite sizable at high energy linear
electron colliders, is also optionally modelled in \kkmceeFive\ using \foam;
see Fig.~\ref{fig:BST}.
In the present version, the BST spectrum is parametrized
as in the {\tt CIRCE1} package in terms of $z_i=E_i/\bar{E}_i$ variables:
\begin{equation}
P(z_1,z_2)=\big[p_0\delta(1-z_1) +p_1 z_1^{p_2} (1-z_1)^{p_3}\big]
\big[p_0\delta(1-z_2) +p_1 z_2^{p_2} (1-z_2)^{p_3}\big].    
\label{eq:BST}
\end{equation}
The parameters $p_i$ of {\tt CIRCE1} have to be provided ``manually''
by the user through input parameters.%
\footnote{Some default values are provided.}
Since \foam\ cannot deal directly with the $\delta(1-z_i)$ distributions,
they are replaced in our \foam\ integrand with very narrow Gaussian peaks.
It is planned to interface \kkmcee\ 5.x with the newer
{\tt CIRCE2} package, which does not have $\delta(1-z_i)$ components.

Thanks to the flexibility of \foam, generating BES and BST simultaneously would not be difficult, but the user must provide the combined BES+BST distribution.

\newpage
\section{Reproducing benchmarks}
\label{sec:benchm}
Benchmarks are numerical results from other programs
and/or older versions of the same program, which should be reproduced
as a proof that the newly developed program works correctly.

\subsection{Muon channel benchmarks}
\begin{table}[h]
\centering
\includegraphics[width=1.02\textwidth]{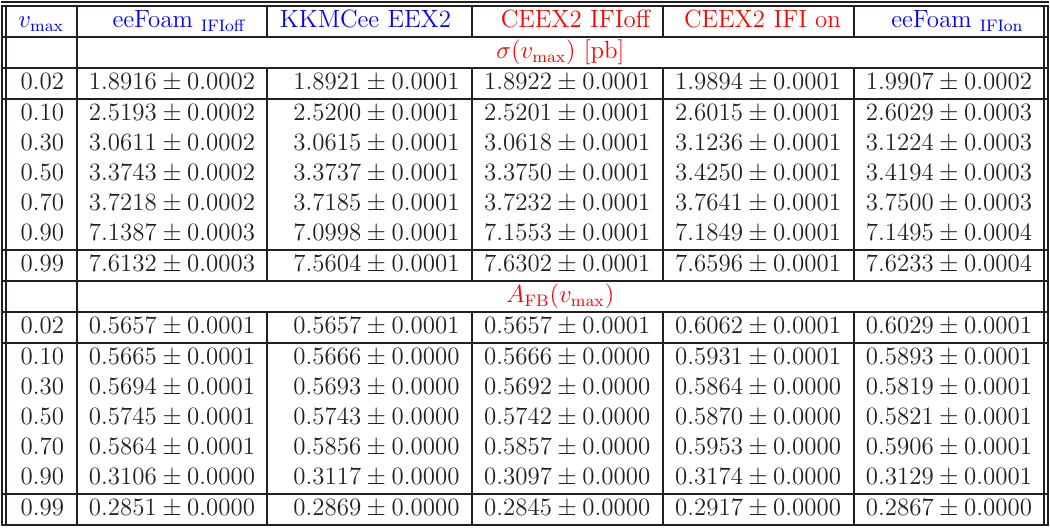}
\caption{
Total cross section and charge asymmetry for the muon pair
production channel at $189$~GeV from \kkmceeFive\ program.
Results are from a run of 700M events with constant MC weights $W=1$.
}
\label{tab:PRD63III}
\end{table}
\begin{table}[!h]
\centering
\includegraphics[width=1.02\textwidth]{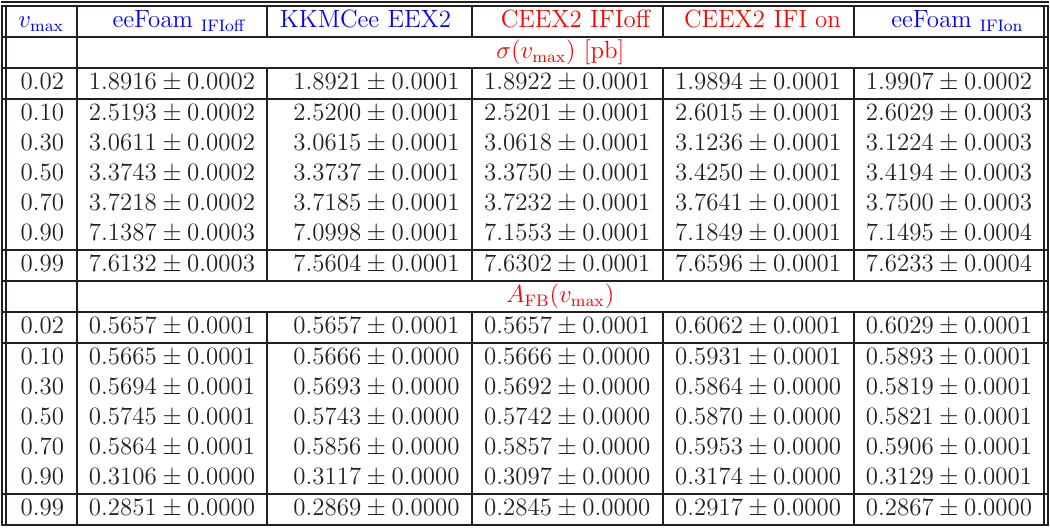}
\caption{
Total cross section  and charge asymmetry for the muon pair
production channel at $189$~GeV from \kkmceeFive.
Results are from a run of 7G variable weight events.
Results from \kkeefoam\ marked as {\tt eeFoam} are also shown.
}
\label{tab:PRD63IIIb}
\end{table}

The standard benchmark for the muon pair production channel,
which was reproduced for all past versions of \kkmc,
was always Table~III of Ref.~\cite{Jadach:2000ir}
for the total cross section $\sigma(v_{\max})$ 
and the charge asymmetry $A_{FB}(v_{\max})$ at 189 GeV.
At this energy  above the $Z$ peak the ``radiative return''
enforces a lot of hard photons, hence higher orders of QED get magnified.
Our new Table~\ref{tab:PRD63III} shows  the cross section  and charge asymmetry
as a function on the total photon energy $v_{\max}$ obtained from \kkmceeFive.
Results are from a 700M events run of $W=1$ events.
The other Table~\ref{tab:PRD63IIIb} shows similar results
from a MC run of 7G variable weight events.
Results from the two runs are compatible within the statistical errors 
and in agreement with Table~III of Ref.~\cite{Jadach:2000ir},
modulo slightly different input parameters 
like masses of $Z$, Higgs boson and $t$ quark.

\newpage
\subsection{Neutrino channel benchmarks}
\begin{table}[!h]
\centering
\includegraphics[width=0.80\textwidth]{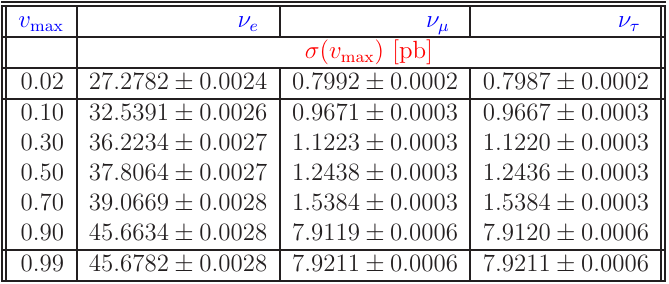}
\caption{
Total cross section for producing the three kinds of SM neutrino pairs
as a function on the cut-off on the total photon energy.
\kkmceeFive \ results are for a 1G event  run of variable weight events.
}
\label{tab:3nunubar}
\end{table}

\begin{figure}[!h]
\centering
\includegraphics[width=0.99\textwidth]{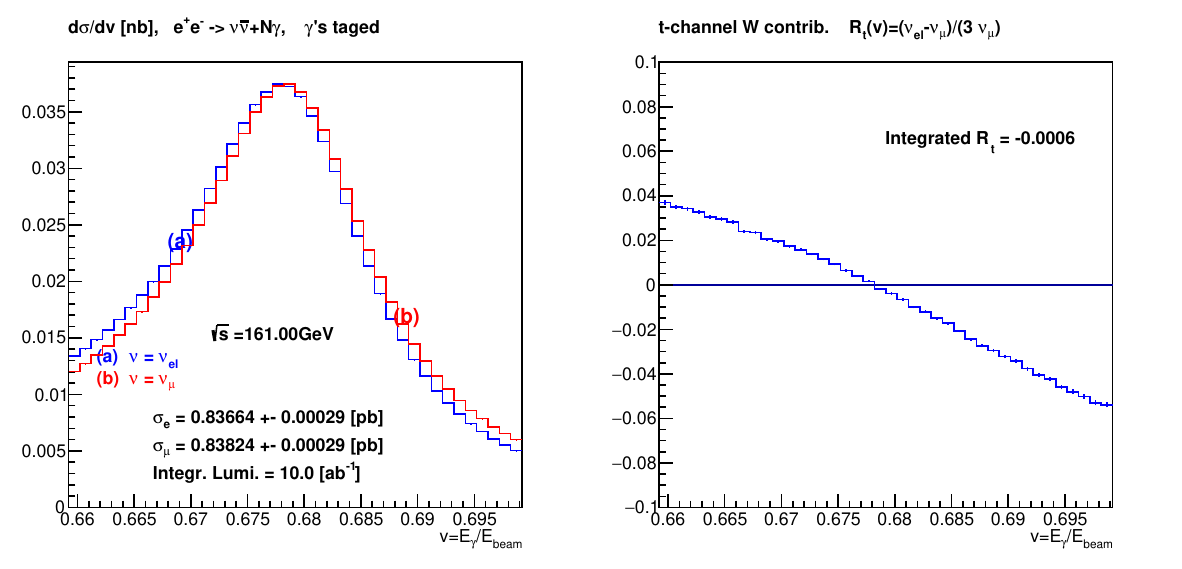}
\caption{
Photon energy distribution for the invisible $Z$ decay.
Cut-offs on photon angle are the same as in ref.~\cite{Aleksan:2019erl}.
Based on a 1G \kkmceeFive\ sample of variable weight events at $161$~GeV.
}
\label{fig:Zinv}
\end{figure}

Table~\ref{tab:3nunubar} shows the total cross section of neutrino pair
production $\sigma_{\nu\bar\nu}$ for $\nu=\nu_{el}$, $\nu_\mu$ and $\nu_\tau$.
There is no other cut-off except the cut-off on the total photon energy 
$v<v_{\max}$, hence this cross section is
``academic'', {\it i.e.} not really observable in an experiment.
Nevertheless, it is worth checking
whether the result for $\sigma_{\nu\bar\nu}(v_{\max})$ from \kkmceeFive\
is the same as that in Table~2 of Ref.~\cite{Bardin:2001vt} obtained 
from the classic F77 version of \kkmc.
New results from \kkmceeFive\ for 1G events presented in Table~\ref{tab:3nunubar}
turn out to be identical with those of Ref.~\cite{Bardin:2001vt},
within the statistical errors.

Another important crosscheck was done by means of reproducing
selected results of Ref.~\cite{Aleksan:2019erl},
with the realistic experimental cutoffs selecting $\nu\bar\nu\gamma$ events.
In our Fig.~\ref{fig:Zinv} we have reproduced the results
of Figure~6 in Ref.~\cite{Aleksan:2019erl}.
It compares the gamma spectrum for the $\nu_{el}$ and $\nu_\mu$ channels.
It is now properly reproduced by  C++ version of \kkmcee.

\subsection{Tau lepton decay benchmarks}
\kkmceeFive\ generates $\tau$ lepton decays using the \tauola\ package,
which is an independent external Monte Carlo event generator.
\tauola\ simulates unpolarized $\tau$ decays in their rest frames.
The important role of \kkmcee\ is to implement spin effects
in $\tau$ decays including spin correlations between the two decays
and to transform decay products from $\tau$ rest frames to the laboratory frame.
Simulating the $\tau$ pair production process, including all QED and EW corrections, 
is performed by \kkmcee\ the same way, of course, as for $\mu$ and quark pairs.
Spin effects are implemented using the algorithm defined in Ref.~\cite{Jadach:1984hwn},
and further developed in Ref.~\cite{Jadach:1998wp}.
Here we include only a few benchmarks relevant for
the implementation of spin effects.
The \tauola\ project has a lot of its own benchmarks, which are briefly
listed in Section \ref{sec:tauola_benchmarks}.

\begin{figure}[!h]
\centering
\includegraphics[width=0.70\textwidth]{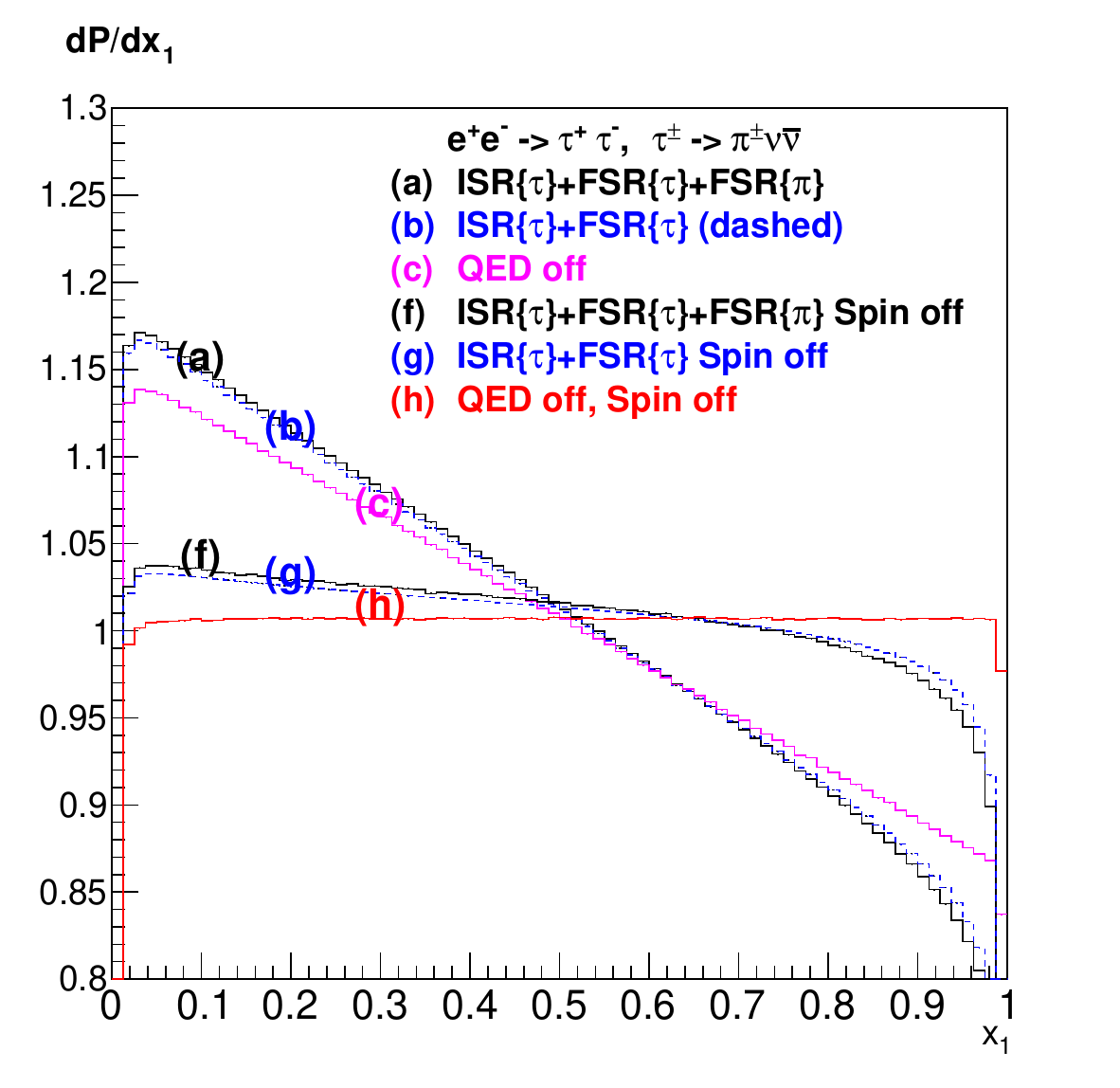}
\caption{
Distributions of the rescaled pion energy variables 
$x_1=E_{\pi^-}/E_{beam}$
with various levels of QED in $\tau$ production and decay,
switching on/off $\tau$ spin polarization effects in the decay.
}
\label{fig:X1pi}
\end{figure}

In the benchmark included in the present distribution directory both $\tau$'s
decay into a single $\pi$ meson, 
$\tau^-\to\pi^-\nu_\tau$, $\tau^+\to\pi^+\bar\nu_\tau$,
at $\sqrt{s}=M_Z$.

Fig.~\ref{fig:X1pi} shows the distribution of the $\pi^-$ energy
from the $\tau^-$ with and without QED in the $\tau$ production process.
All distributions in Fig.~\ref{fig:X1pi} are normalised to one.
The nonzero slope of the pion energy distribution reflects
mainly the nonzero spin polarization of the $\tau$ lepton.
However the pion energy spectrum is also softened by the photon emission
from incoming beams (ISR), outgoing $\tau$'s (FSR)%
\footnote{Initial-final state QED interference (IFI) is
also switched on, but its effect is negligible}
and out of $\pi$ mesons after $\tau$ decays (generated by \photospp).
The QED effects and polarization effects in Fig.~\ref{fig:X1pi} 
are switched on/off in order to see them separately.
The QED effects obtained from \kkmceeFive\ look the same as in Fig.5 
in Ref.\cite{Eberhard:1989ve,Altarelli:1989hv}.
Note that the structures visible in the distribution of Fig.~\ref{fig:X1pi}
close to $x_1=0$ and $x_1=1$
are of kinematic origin due to finite $\tau$ mass
and will look different at other CMS energies.

\begin{figure}[!h]
\centering
\includegraphics[width=0.49\textwidth]{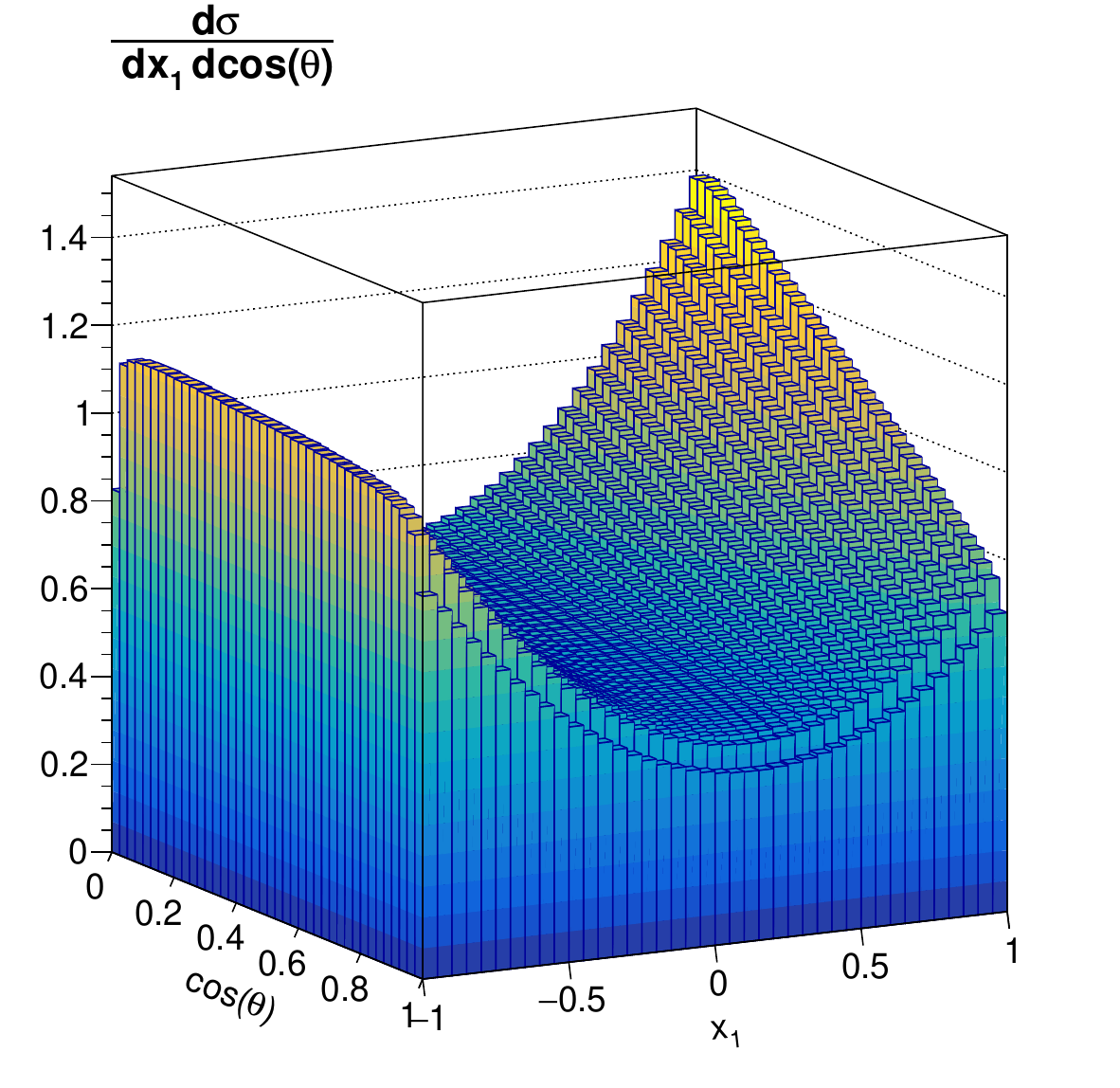}
\includegraphics[width=0.49\textwidth]{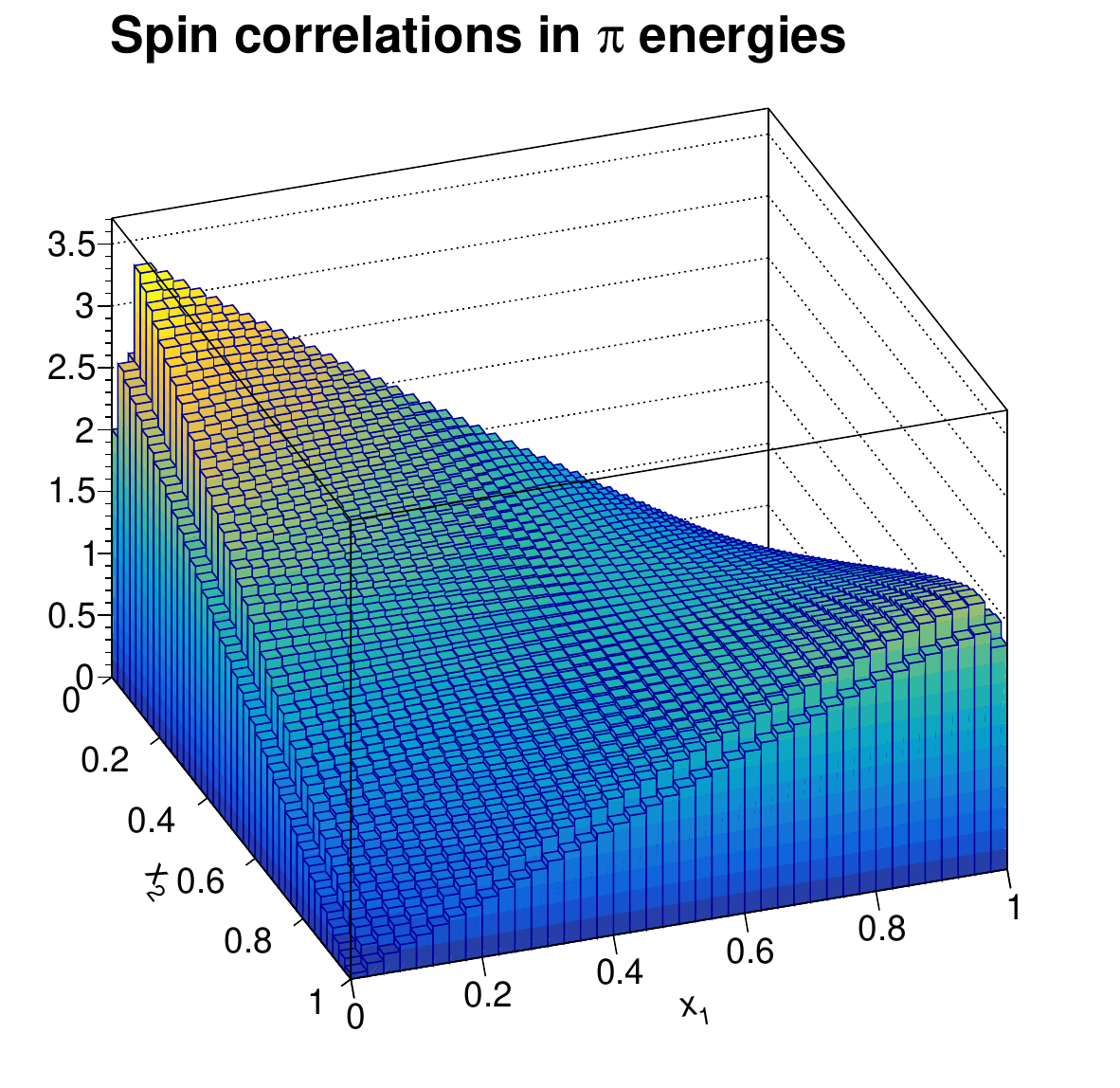}
\caption{
Distributions of the rescaled pion energy variables 
$x_1=E_{\pi^-}/E_{beam}$, $x_2=E_{\pi^+}/E_{beam}$
and the polar angular variable $\cos(\theta_{\pi^-})$ of the $\pi^-$ obtained from \kkmceeFive.
}
\label{fig:taupair2}
\end{figure}

The left panel of Fig.~\ref{fig:taupair2} shows the double differential distribution
of the rescaled energy $x_1=E_{\pi^-}/E_{beam}$
and the angular variable $\cos(\theta_{\pi^-})$ of the $\pi^-$ meson
originating from the decayed $\tau^-$.
We see clearly that the $\pi^-$ energy distribution varies strongly with
the $\pi^-$ ($\tau^-$) angle, reflecting the fact that the $\tau^-$
spin polarization depends on the $\tau^-$ polar angle.
This plot is essentially the same as Fig.~1 
in Ref.\cite{Eberhard:1989ve,Altarelli:1989hv} obtained using the classic 
F77 version of \kkmc.

On the other hand, the right panel of Fig.~\ref{fig:taupair2} 
illustrates the well-known phenomenon of strong spin correlations
among the two $\tau$ leptons, which are seen in the double energy distribution
of the two $\pi$ mesons from the two decaying $\tau$'s.

\subsection{Benchmarks of \tauola}
\label{sec:tauola_benchmarks}
The $\tau$-lepton decay library \tauola\  
is a separate project independent of \kkmcee\ with its own benchmarks,
which are expected to undergo substantial upgrades in the near future,
mainly due to the impact from the new Belle 2 data.
For the convenience of the reader let us list
benchmarks of \tauola, even if they
are strictly-speaking outside of the scope of the present paper.
They can be divided into the following groups:
\begin{enumerate}
\item[A] Basic functionality tests.
\item[B] Tests of $\tau$ decay initialization.
\item[C] Tests of event record content.
\item[D] Tests of longitudinal  spin correlations.
\item[E] Test of polarization as a function of phase space regions.
\item[F] Tests of transverse spin correlations.
\end{enumerate}

Tests of the group [A] are explained in Ref.~\cite{Kaczmarska:2014eoa};
they monitor longitudinal polarization and spin correlation effects transmitted 
to the simplest tau decays:
$\pi^\pm \nu_\tau$, 
$\pi^\pm \pi^0\nu_\tau$, 
$l^\pm \nu_\tau \bar\nu_l$, 
Typically, the ratio of the decay product energy to the energy of decaying $\tau$-lepton 
is used. For the longitudinal spin correlations, the ratio of the invariant mass of 
the visible tau lepton pair decay products to the lepton pair invariant mass is used. 
One dimensional histograms or two dimensional scattergrams are used. 
The $Z$ peak region is used for the center of mass energy.
These tests are not new and they were already used in \cite{Eberhard:1989ve}.
They were continuously used in one form or another for many years. 

These tests are also available with the help of MC-tester~\cite{Davidson:2008ma}. 
The purpose of this tool is to monitor content of the generated event records by 
different programs and in different languages. 
That is why MC-tester is useful for the tests of group [B] and [C]. 
From the generated sample event tree the branch starting from $\tau^+$ 
or $\tau^-$ is identified. If it represents a new decay channel,
it automatically defines histograms of all invariant masses
which can be constructed from stable final state decay products\footnote{In this case, stable final decay products of $\tau^+$ (or $\tau^-$).}.
At the end, the list of  decay channels is created and 
all may be compared with the results from another run,
possibly of another program. 
In this way, the correctness of the decay channel choices and
frequencies of their occurrences can be stored. 
If any of the decay channels is improperly stored in the event record, 
there is a good chance that this will be exposed with the run.

For test groups [D] and [E], when monitoring invariant masses, one has to monitor 
the system decaying to the $\tau$-lepton pair down to the decay products of both taus.
That is why MC-tester can be of help in these cases too. 
For the test, one can select samples from distinct phase space regions, 
not necessarily close to the $Z$ peak, as in Ref.~\cite{Kaczmarska:2014eoa}. 
The impact of the electroweak initialisation on observables sensitive to $\tau$ polarization 
can be studied in this way as well.

Observables sensitive to transverse spin cannot be studied with one-dimensional invariant mass distributions. 
For such tests, one can use distributions of the acoplanarity angle,
as for CP-sensitive observables in Higgs studies \cite{Jozefowicz:2016kvz}.
However, in the case of a $\tau$-pair produced from a 
vector/pseudovector state, additional cuts are needed.
We think that old tests of low energies~\cite{Jadach:1984hwn} (eq. 2.6 there) 
remain the best until now. The advantage is that $\tau$ mass terms, 
important for low effective mass $\tau$ pairs, are addressed for these old tests.

\section{Summary and outlook}
\label{sec:prospects}

In this work, we present a new version of \kkmcee\ written entirely in C++,
except two external libraries (\dizet\ and \tauola) which temporarily remain in F77.
Functionally and from the point of view of physics content, the new program inherits all essential features of the previous F77 version.
However, a number of important improvements are introduced.
The present document does not repeat that which has not changed
from the f77 version and is thoroughly covered in Ref.~\cite{Jadach:1999vf}.
On the other hand, we make quite some effort to describe 
and reproduce many high precision benchmarks of \kkmcee, 
that is, the kind of tests which show that the present C++ version
correctly reproduces results obtained in many past papers
(using older F77 versions), 
for all final state channels and for many observables.
The user may easily reproduce/repeat these benchmarks
and compare them with tables and plots in this paper,
or with archived outputs included in the distribution directory.

The most important aim of the C++ version is to provide a flexible basis
for the future development of \kkmcee.
Let us list at least some of them:
\begin{itemize}
\item 
Adding CEEX \order{\alpha^n L^n}, n=3, corrections,
maybe also for $n=4,...,\infty$, while maintaining the soft limit.
\item 
Forcing a visible photon at the generator level in the neutrino pair channels.
\item 
Automated construction of the CEEX matrix element, 
for porting it to other processes like $HZ$ production and decay.
\item
A new \order{\alpha^2} EW library. 
Note, however, that 1-loop electroweak corrections 
for the differential distributions for
$e^+e^-\to f\bar{f}\gamma$ are still missing in the literature%
\footnote{Only those averaged over the real photon were calculated.}!
\item
Making the MC algorithm for the multi-photon
phase space more efficient in some corners of the phase space 
(2 very hard photons).
\item
Integrating the Bhabha process into \kkmcee? It is thinkable, 
but only provided a good quality EW library (in C++) is available.
\end{itemize}
There are, of course, many other interesting development avenues as well.

\vfill
\newpage
{\bf\Large Appendix}
\appendix
\section{Monte Carlo weight distribution}
\label{apx:mc-weight}

Using the steering parameter {\tt KeyWGT}, the user may decide
to generate  events with variable weight $W$ or constant $W=1$ events.
Weighted events are better suited for theoretical studies, while
$W=1$ events are often mandatory for the experimental studies and data analysis. 
This is because the detector simulation of a single MC event
usually requires much more CPU time than the generation 
of the MC event using our program.
The MC weight of \kkmcee\ is normalized such that its average
$\Aver{W}$ is not far from one.
$W=1$ events are obtained from the variable weight events 
using the standard rejection method,
that is accepting events with $W< r W_{\max}$, where $r$ is random number,
and trashing the MC event otherwise.
The default value $W_{\max}=4$ is used in our program
and it can be redefined by the user through one of the input parameters.
The acceptance rate is roughly $\Aver{W}/W_{\max}$.
If the MC weight is very regular, without a tail, then $W_{\max}$
can be adjusted to get acceptance not far from 100\%.
In our case the MC usually has some tail dependent on the process type,
CMS energy, dummy IR cutoff $v_{\min}$, 
maximum allowed photon energy parameter $v_{\max}$, etc.
In the following we shall elaborate on these dependencies.

The typical weight distribution of \kkmceeFive\ is shown in the upper panel of Fig.~\ref{fig:A0_proposal}.
The main weight of the second order CEEX has a tail extending up to $W=4$.
In the case of IFI switched off the weight tail extends only up to 2.
FOAM does not contribute to the main MC weight,
because in the present version, FOAM provides $W=1$ events 
(this can be changed using input parameters).
The above shape of the weight distribution will be reflected
in the following numerical analysis.
Curve (B) for the weight with IFI switched off has 
some multi-peak structure, which comes from the low level FSR generator. 
This structure is smeared out by the IFI component of the weight, if present.

In the classic version of \kkmc, the modelling of the lepton $\cos\theta$
distribution was included in the main weight, while in the present
\kkmceeFive\ version the dependence on $\cos\theta$
is modelled by \foam.
The main MC weight distribution of the classic \kkmc\ with the flat
baseline angular distribution of the final fermions,
in the presence of IFI, has a much longer tail than does 
that of the C++ version of \kkmcee, extending beyond $W=8$. 

Let us explain now how we advise the user to deal
with typical cases where the fraction of
the potentially harmful MC events with $W>W_{\max}$ is not completely negligible%
\footnote{This method is also used in \foam\ since a long time ago.}.
This is done such that for accepted MC events with $W< r W_{\max}$ 
we assign a new weight $W'=1$ while for accepted MC events with $W> W_{\max}$
we assign a new weight $W'=W/W_{\max}>1$.
The user should record this new weight and use it for controlling
and mitigating unwanted effects.
First of all, if this new weight is used in all distributions,
then one gets exactly the same result as for variable weight events,
without any bias depending on $W_{\max}$.
If the $W'$ weight is neglected then one is back in the standard rejection method
with possible bias due to ``mistreating'' $W> W_{\max}$ events.
However, one may easily evaluate how big is this bias using additional
histograms with the $W'-1$, see below.
This makes sense because in the sub-sample of the events which are
of the main interest of the user, the effect due to $W'-1$
may be much smaller (or bigger) than for the entire MC sample.

Also, there is a possibility for events stored on the disk to apply 
another secondary rejection procedure for $W'$ using a new $W'_{\max}$,
instead of generating a new series of MC events with the higher $W_{\max}$.
However, one has to be careful in such a scenario with keeping
track of the overall normalization.
Obviously, in any case it is worth while to preserve $W'$ in the event record.

\begin{figure}[!h]
\centering
\includegraphics[width=0.80\textwidth]{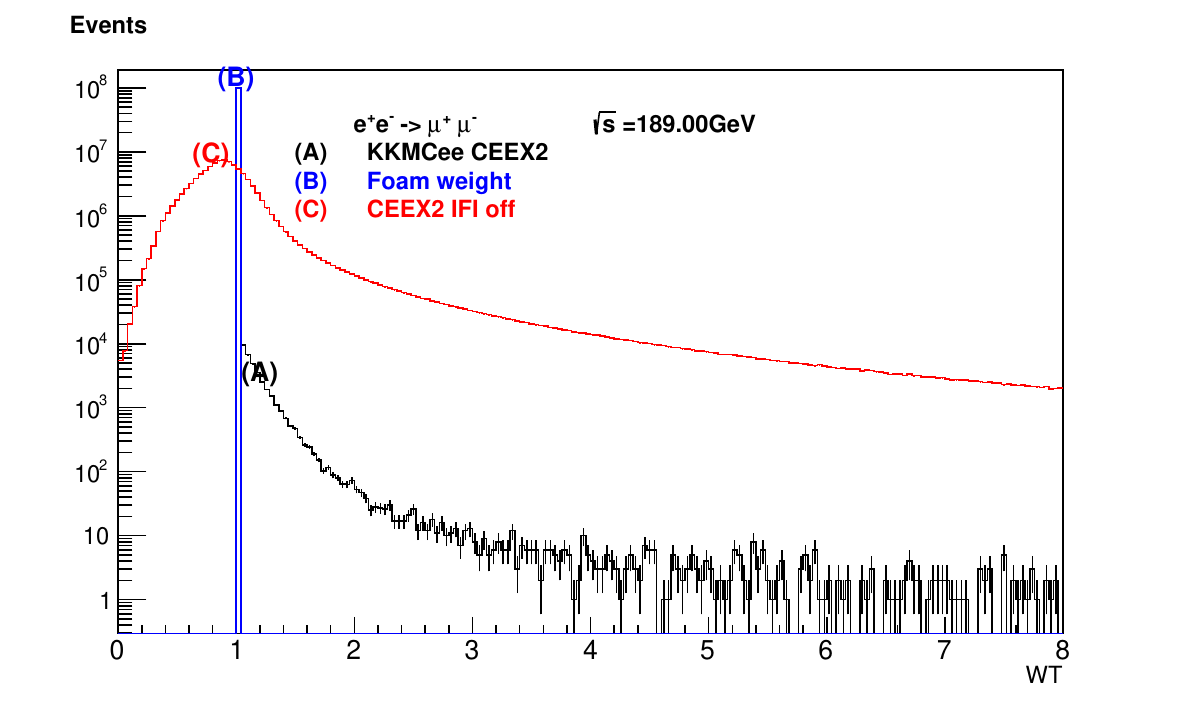}
\caption{
Distribution of the weight $W'$ for the muon pair channel at 189 GeV
for 18M events. In most events, $W'=1$, but spillover events 
with $W'$ are visible.
}
\label{fig:A02}
\end{figure}

In Fig.~\ref{fig:A02}, we see the example of the distribution of $W'$
with a $W'>1$ spillover tail at the $10^{-4}$ level.
Since the main weight including IFI is made equal one, 
the auxiliary weight undoing the IFI effect, which is also 
shown in Fig.~\ref{fig:A02}, has a quite wide shape.

\begin{table}[!h]
\centering
\begin{tabular}{|l|l|l|l|l|l|}
\hline
\multicolumn{6}{|c|}{$\mu+\tau$, ISR+FSR+IFI, $W_{\max}=4$} \\ \hline
\hline 
$\sqrt{s}$ & $\Aver{W}$ & $\sigma(W)$ & $R_{acc}$ & $\Delta_{over}$ & $W_{sup}$ \\
\hline 
189GeV  & 0.2900  & 0.3700  & 0.0720 & 1.67$\times 10^{-5}$ & 4.676 \\ 
\hline 
161GeV  & 0.3181  & 0.3754  & 0.0805 & 5.33$\times 10^{-5}$ & 4.851 \\ 
\hline 
105GeV  & 0.5076  & 0.4467  & 0.1296 & 2.08$\times 10^{-5}$ & 4.815 \\ 
\hline
94.3GeV & 0.7213  & 0.5304  & 0.1812 & 1.05$\times 10^{-5}$ & 4.406 \\ 
\hline 
91.2GeV & 0.8816  & 0.6076  & 0.2209 & 6.07$\times 10^{-5}$ & 4.975 \\ 
\hline
87.9GeV & 0.7945  & 0.5479  & 0.2010 & 12.7$\times 10^{-5}$ & 5.837 \\ 
\hline 
10.0GeV & 0.4921  & 0.4921  & 0.1227 & 0.000                & 3.903 \\
\hline
\end{tabular}
\caption{
Results on several parameters
of the MC weight (CEEX) for several CMS energies
obtained from a MC run done for $10^4$ accepted $W=1$ events.
}
\label{tab:A1}
\end{table}

Table~\ref{tab:A1} shows results on several parameters
of the MC weight (CEEX) for several CMS energies
obtained from the MC run done for $10^4$ accepted $W=1$ events.
If not stated otherwise then the IR cut-off was $v_{\min}=10^{-5}$ 
and the upper photon phase space limit was $v_{\max}=0.99$.
Listed in this table are:
average MC weight $\Aver{W}$ and its dispersion $\sigma(W)$,
fraction $R_{\rm }$ of accepted MC events,
fraction of over-weighted cross section defined as
$\Delta_{\rm over}= \Aver{W-W_{\max}}/\Aver{W}$
and  $W_{\rm sup}$, the biggest MC weight in the run.
All results are in Tab.~\ref{tab:A1}.
As we see from the table, in some cases one could lower 
$W_{\max}$ from 4 down to 3 without significant increase of $\Delta_{\rm over}$.

\begin{table}[!h]
\centering
\begin{tabular}{|l|l|l|l|l|l|}
\hline
\multicolumn{6}{|c|}{$\mu+\tau$, ISR+FSR, IFIoff, $W_{\max}=2$} \\ \hline
\hline 
$\sqrt{s}$ & $\Aver{W}$ & $\sigma(W)$ & $R_{acc}$ & $\Delta_{over}$ & $W_{sup}$ \\
\hline 
189GeV & 0.5105  & 0.4087  & 0.2545 & 0.0000 & 1.759 \\ 
\hline 
105GeV & 0.8402  & 0.4300  & 0.4208 & 0.0000 & 1.844 \\ 
\hline
91.2GeV& 1.1623  & 0.4100  & 0.5803 & 0.0000 & 1.594 \\ 
\hline
10.0GeV& 0.7235  & 0.4856  & 0.3629 & 0.0000 & 1.383 \\
\hline
\end{tabular}
\caption{
Results on several parameters
of the MC weight (CEEX) for several CMS energies.
IFI is now switched off.
}
\label{tab:A2}
\end{table}

Table~\ref{tab:A2} shows similar results for
the muon and tau pair production, but switching off the initial-final
state interference IFI.
As expected, the weight distribution is much better, 
with the zero contribution from over-weighted events above $W_{\max}=2$
and a much higher acceptance rate.

\begin{table}[!h]
\centering
\begin{tabular}{|l|l|l|l|l|l|}
\hline
\multicolumn{6}{|c|}{$\nu_\mu$, ISR, $W_{\max}=4$} \\ \hline \hline 
$\sqrt{s}$ & $\Aver{W}$ & $\sigma(W)$ & $R_{acc}$ & $\Delta_{over}$ 
& $W_{sup}$ \\ \hline 
189GeV & 0.4199  & 0.3223 & 0.1064 & 0.000 & 1.351 \\ 
\hline 
161GeV & 0.4876  & 0.3345 & 0.1218 & 0.000 & 1.333 \\ 
\hline
105GeV & 0.8698  & 0.2766 & 0.2170 & 0.000 & 1.295 \\ 
\hline \hline 
\multicolumn{6}{|c|}{$\nu_{e}$, ISR, $W_{\max}=4$} \\ \hline
189GeV & 1.4405  & 1.3443 & 0.3582 & 0.000 & 3.806 \\ 
\hline 
161GeV & 1.2873  & 1.2293 & 0.3214 & 7$\times 10^{-7}$ & 4.0886 \\ 
\hline
105GeV & 1.1241  & 0.5759 & 0.2814 & 0.0000 & 2.572 \\ 
\hline \hline 
\end{tabular}
\caption{
Results on several parameters
of the MC weight (CEEX) for several CMS energies
for $\nu_\mu \bar\nu_\mu $ and $\nu_{e} \bar\nu_{el}$ production, for 50k events.
}
\label{tab:A3}
\end{table}

Let us now switch to neutrino final states.
In Tab.~\ref{tab:A3} we show similar results as in the previous tables, 
separately for the muon neutrino and for the electron neutrino.
For the muon neutrino the weight parameters are quite similar 
as for the muon pair without IFI. 
The common $W_{\max}=4$ is satisfactory for the rejection method
providing $W=1$ events.
For the electron neutrino, the weight distribution is also quite satisfactory, 
in spite of the presence of the $t$-channel $W$ exchange 
peak in the neutrino angular distribution.
However, in the classic \kkmc, this $t$-channel peak electron neutrino was causing
a $50\%$ contribution of overweighted events
even for $W_{\max}=16$, because of a long tail in
the weight distribution was extending to $W=200$.
Thanks to use of \foam\ for the angular distribution of the neutrino,
this bad behaviour of the main weight is now cured completely!

\begin{table}[!h]
\centering
\begin{tabular}{|l|l|l|l|l|l|}
\hline
\multicolumn{6}{|c|}{$\mu+\tau$, ISR+FSR+IFI } \\ \hline
\hline 
$\sqrt{s}$ & $\Aver{W}$ & $\sigma(W)$ & $R_{acc}$ & $\Delta_{over}$ & $W_{sup}$ \\
\hline 
189GeV, $W_{\max}=4$ & 0.2900  & 0.3700  & 0.0720 & 1.67$\times 10^{-5}$ & 4.676 \\ 
\hline
189GeV, $W_{\max}=2$ & 0.2908  & 0.3700  & 0.1434 & 0.0078 & 3.956 \\ 
\hline\hline
\end{tabular}
\caption{
Results on several parameters
of the MC weight (CEEX) for several CMS energies
obtained from MC run done for $10^4$ accepted $W=1$ events.
}
\label{tab:A4}
\end{table}

Table~\ref{tab:A4} illustrates what happens when the value of
$W_{\max}$ set for $W=1$ events is too small.
Reducing $W_{\max}$ by factor 2 improves acceptance rate by factor 2,
but the contribution of overweight events jumps to 0.7\%, 
which may be considered as not acceptable.

\begin{figure}!h]
\centering
\includegraphics[width=0.49\textwidth]{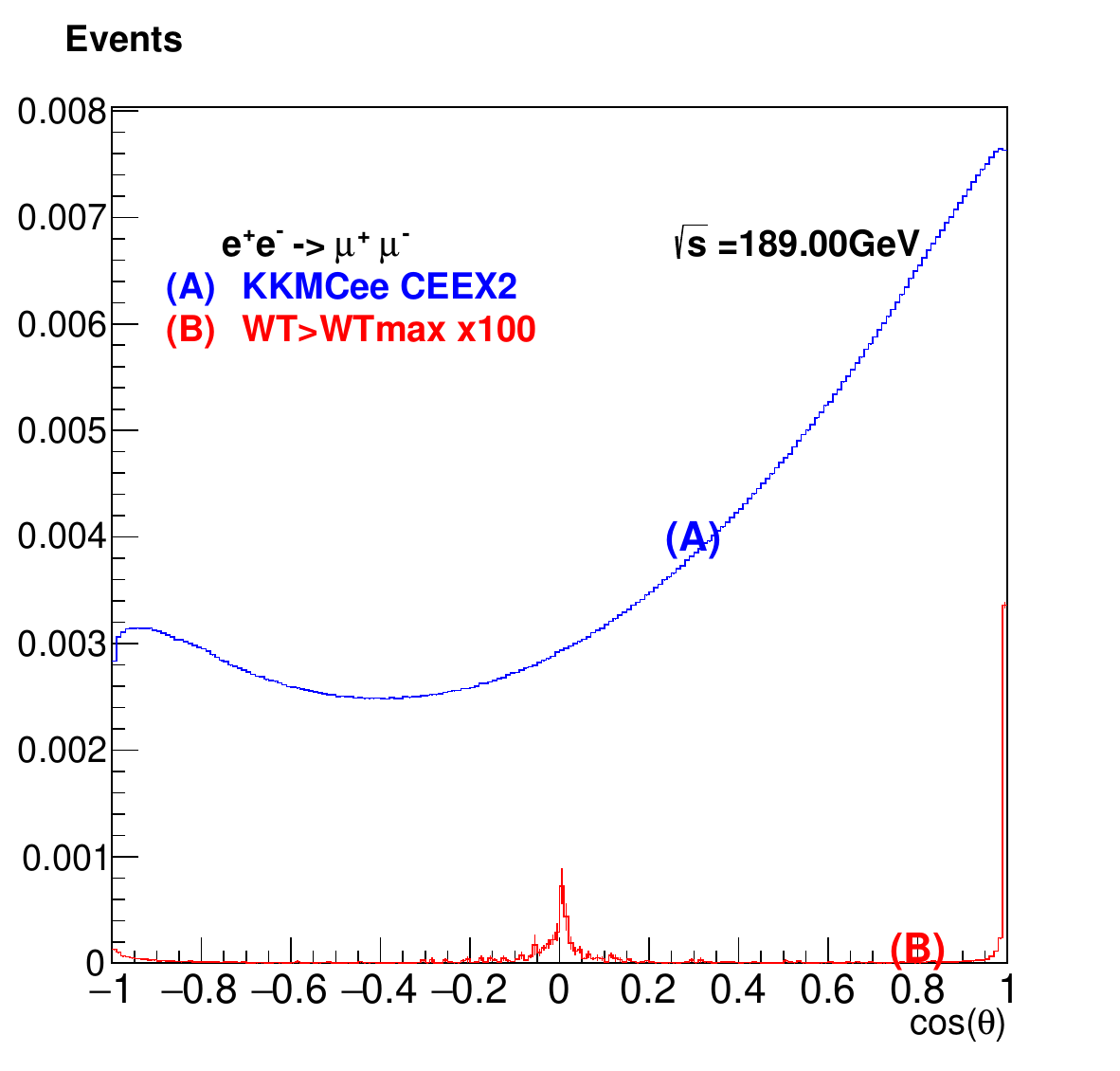}
\caption{
The angular distribution of leptons is shown for the entire
MC sample with $v<v_{\max}=0.99$ together with the contribution 
$W'>1$ events.
The excess $W'-1$ weight is also recorded.
Results are for the default $W_{\max}=4$.
The contribution from $W'>1$ events
is magnified by factor 100 to be visible.
}
\label{fig:A5}
\end{figure}

In a realistic experiment, the angular range of the lepton is often limited
to $|\cos(\theta)|<0.9$. As we have already indicated, 
the ``bad'' events with $W'>1$ spoiling $\Delta_{over}$
are expected to concentrate close to $\cos(\theta)=1$
(due to well known IFI properties) 
and they should be practically irrelevant elsewhere.
Fig.~\ref{fig:A5} illustrates this phenomenon quite clearly.

\vspace{10mm}
\noindent
\section*{Acknowledgments}

The authors would like to thank Marcin Chrz\c{a}szcz and Jacek Holeczek 
for help in maintaining and improving the source code of the program.


\end{document}